\documentclass[12pt, letterpaper]{article}
\usepackage{amssymb,amsfonts,amsmath}
\usepackage{setspace}
\usepackage[top=25mm, bottom=25mm, left=25mm, right=25mm]{geometry}

\usepackage{graphicx}
\usepackage{natbib}

\newcommand{\gene}[1]{{\it #1}}
\newcommand{\EQ}[1]{Equation~(\ref{eq:#1})}
\newcommand{\FIG}[1]{Figure~\ref{fig:#1}}

\newcommand{\mut}{u}
\newcommand{\nref}{\hat{n}}
\newcommand{\is}{i_c}
\newcommand{\ns}{n_c}
\newcommand{\xs}{x_c}
\newcommand{\gr}{g}
\newcommand{\su}{\tilde{s}}
\newcommand{\sur}{\tilde{s}}

\begin{document}


\title{Estimating the Strength of Selective Sweeps from Deep Population Diversity Data}
\author{Philipp W.~Messer${}^\dagger$ and Richard A.~Neher${}^\ast$\vspace{.5cm}\\
${}^\dagger$ Department of Biology, Stanford University, Stanford, California 94305, USA\\ ${}^\ast$ Max-Planck-Institute for Developmental Biology,
72076 T\"ubingen, Germany }
\date{\today}

\maketitle

\vspace{1cm}
\noindent key words: selective sweep, deep diversity data, selection coefficient

\noindent running title: Estimating the Strength of Selective Sweeps

\vspace{1cm}
\noindent Corresponding author: \\
Richard Neher \\
MPI for Developmental Biology \\
Spemannstr.~35\\
72076 T\"ubingen \\
Germany \\
Tel: +49 (0)7071 601 1345 \\
email: richard.neher@tuebingen.mpg.de

\newpage

\section*{Abstract}

Selective sweeps are typically associated with a local reduction of genetic diversity around the adaptive site. However, selective sweeps can also quickly carry neutral mutations to observable population frequencies if they arise early in a sweep and hitchhike with the adaptive allele. We show that the interplay between mutation and exponential amplification through hitchhiking results in a characteristic frequency spectrum of the resulting novel haplotype variation that depends only on the ratio of the mutation rate and the selection coefficient of the sweep. Based on this result, we develop an estimator for the selection coefficient driving a sweep. Since this estimator utilizes the novel variation arising from mutations during a sweep, it does not rely on preexisting variation and can also be applied to loci that lack recombination. Compared with standard approaches that infer selection coefficients from the size of dips in genetic diversity around the adaptive site, our estimator requires much shorter sequences but sampled at high population depth in order to capture low-frequency variants; given such data, it consistently outperforms standard approaches. We investigate analytically and numerically how the accuracy of our estimator is affected by the decay of the sweep pattern over time as a consequence of random genetic drift and discuss potential effects of recombination, soft sweeps, and demography. As an example for its use, we apply our estimator to deep sequencing data from HIV populations.
\newpage

\section*{Introduction}

The frequency and strength of positive selection are key parameters of the evolutionary process yet reliable estimates are often very difficult to obtain~\citep{Nielsen:2005p42523,EyreWalker:2006p9603}. As it becomes increasingly practicable to analyze the genetic diversity of natural and experimental populations at high depth, we can hope to obtain better estimates from a detailed analysis of the signatures positive selection is expected to leave in such data.

The standard model describing the population genetic signatures of positive selection is the so-called selective sweep~\citep{Smith:1974p34217}.
Selective sweeps produce distinct effects on the patterns of genetic diversity in the vicinity of the adaptive site: (i) Hitchhiking of linked neutral polymorphism with the adaptive allele generates dips in diversity that level off with increasing genetic distance from the selected site~\citep{Smith:1974p34217}. (ii) Because different lineages from a population sample taken after the sweep can coalesce almost instantaneously during the early phase of the sweep, there should be an excess of singletons in the neutral diversity around the selected site~\citep{Slatkin:1991p43283,Barton:1998p28270,Durrett:2004p41537}. (iii) Derived alleles can be brought to high frequencies~\citep{Fay:2000p35077}, which is unlikely under random genetic drift alone. (iv) The adaptive haplotype should be longer than expected under neutrality since recombination has had only a short time to degrade it~\citep{Hudson:1994p43479}. These hallmark signatures underlie the majority of approaches to scan for recent adaptation in population genetic data~\citep{Hudson:1987p35717,Tajima:1989p37563,Wiehe:1993p37333,Fay:2000p35077,Sabeti:2002p18853,Kim:2002p37296,Przeworski:2003p41514,Nielsen:2005p42523, Voight:2006p18353,Sabeti:2007p12131,Andolfatto:2007p18246}.

In addition to detecting selective sweeps, one would often like to know the strength of selection that drove the sweep~\citep{Macpherson:2007p18421,Hernandez:2011p39635,Sattath:2011p42514}. One common approach is to infer selection coefficients $s$ of the adaptive allele from the size of the dip of approximate length $s/r$ around the adaptive site, where $r$ is the rate of recombination per length in the corresponding region~\citep{Smith:1974p34217,Kaplan:1989p34931,Kim:2002p37296}. Approaches based on this signature rely on the interplay between recombination and ancestral variation and are thus not applicable for scenarios where recombination or ancestral variation is poorly characterized. Furthermore, certain scenarios of adaptation do not always generate clear dips in diversity. Examples are incomplete sweeps where the adaptive mutation is not fixed in the population, and soft sweeps, where more than one haplotype has swept through the population~\citep{Pritchard:2010p35368}.

Here, we develop an estimator for selection coefficients at candidate loci where a selective sweep has occurred recently. Our estimator is based on the analysis of the novel haplotype variation that arises from neutral mutations on the sweeping adaptive haplotype. These early variants are very different in kind from the variation due to neutral mutations occurring after a sweep, since they have experienced an initial phase of exponential amplification. Mutations after the sweep will quickly outnumber the few that happened early during the sweep, but they will drift to higher population frequencies comparatively slowly.

\begin{figure}[h!]
\begin{center}
\includegraphics[width=0.48\columnwidth]{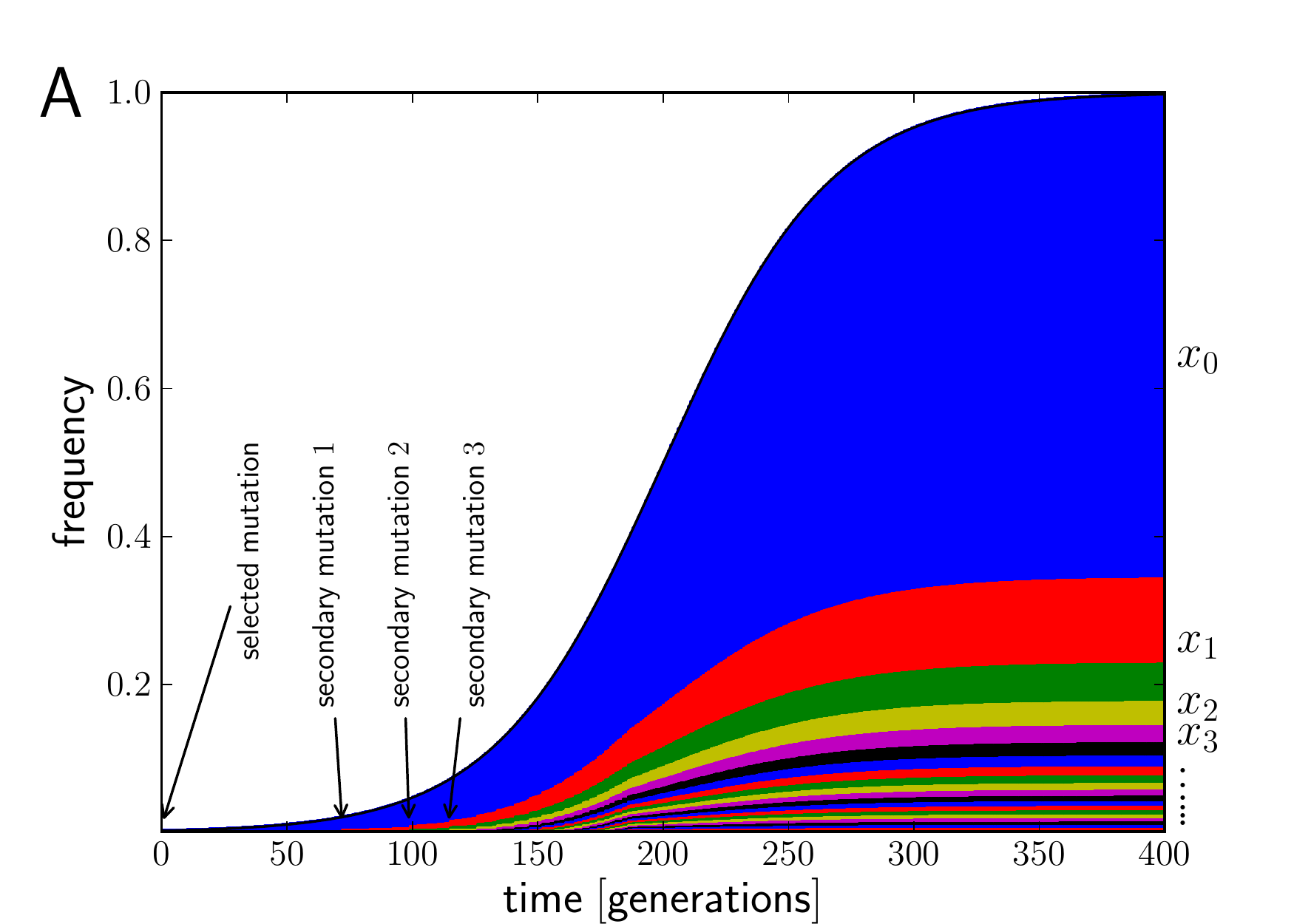}
\includegraphics[width=0.48\columnwidth]{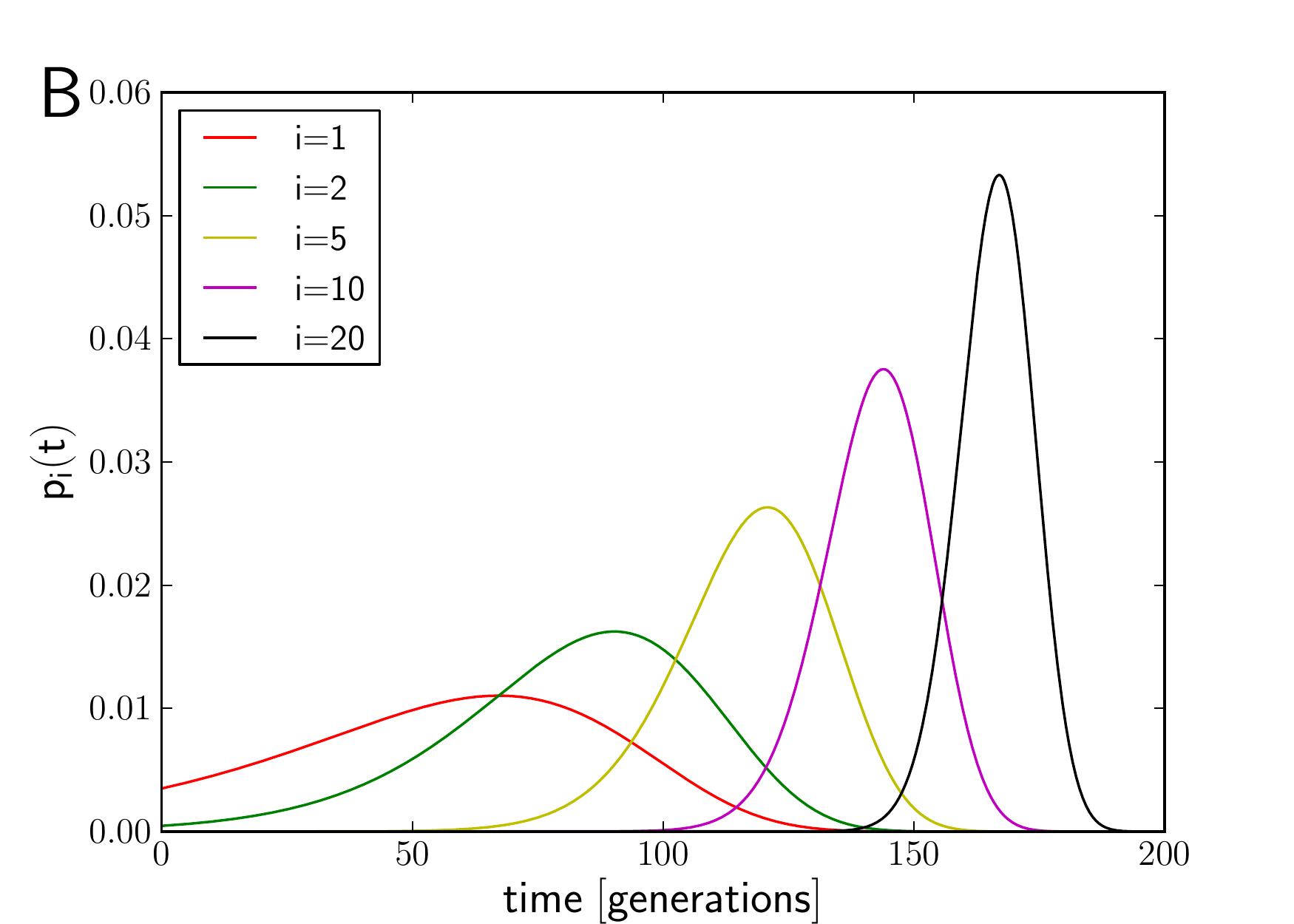}
\caption[labelInTOC]{(A) Generation of novel haplotype diversity from secondary mutations during a selective sweep. At $t=0$ a beneficial mutation establishes and the frequency, $x_0$, of its underlying haplotype (blue) rises. After some time, a neutral mutation occurs in an individual carrying the beneficial mutation, creating a new adaptive haplotype variant (red) which itself increases in frequency, $x_1$. This happens repeatedly, giving rise to a range of low-frequency haplotypes that all descended from the initial founding haplotype. Sweep parameters are $s=0.03$ and $u=0.004$, seeding times of new variants are given by \EQ{origtime}. (B) The distribution of seeding times for different $i$ according to \EQ{dis_of_t}. }
\label{fig:sweep_illustration}
\end{center}
\end{figure}

The frequency spectrum of the early haplotype variants is determined by the distribution of their seeding times during the sweep, as illustrated in \FIG{sweep_illustration}. We show analytically that their rank-frequency spectrum (the frequencies ordered by decreasing abundance) decays according to a simple power-law, which differs markedly from the approximately exponential decay expected under neutral evolution. This power-law is characterized by only a single parameter: the ratio of the rate at which new haplotypes arise and the selection coefficient $s$ of the sweep.

We use this result to develop an estimator of $s$ that compares the strength of selection to the rate at which novel haplotypes are produced. Novel haplotypes can be produced by mutation or recombination. In many organisms, recombination is rare and the mutation rate is larger or similar to the recombination rate~\citep{Roach:2010p34343,Ossowski:2010p32056,HaagLiautard:2007p46053}. Hence our estimate is much less sensitive to poorly characterized recombination rates, thereby overcoming several problems of estimators based on dips in diversity. In particular, our estimator should also be applicable to organisms that exchange genomic segments via horizontal transfer (many bacteria), and populations where levels of ancestral variation are very low, for instance in experimental evolution with clonal populations.

Our estimator relies on deep diversity data for accurate measurements of the population frequencies of rare haplotypes. With large-scale sequencing projects such as the 1000 genomes project in humans~\citep{1000GenomesProjectConsortium:2010p37747} and similar projects in flies~\citep{DPGP} and plants~\citep{Cao:2011p43483} currently under way, such data will become available soon.

\section*{Materials and Methods}

\subsection*{Forward simulations of selective sweeps}

Simulations were performed with custom written C++ programs that represent each haplotype in the population by a bit string of fixed length $L=1024$ or $L=4096$. Populations are initialized with a sample of size $10^4$ individuals obtained from the neutral coalescent using the program \texttt{ms}~\citep{Hudson:2002p37266}. Note that this initial variation is only needed to allow diversification by recombination. We constrained \texttt{ms} to return a sample with $1023$ (or $4095$, respectively) segregating sites, leaving the site in the center of the locus for the beneficial mutation. Each genotype sampled from the coalescent is copied $N_0/10^4$ times into the initial population. The beneficial mutation is introduced into one randomly chosen individual in generation 0. If the beneficial mutation is lost due to random drift, the population is reset to the neutral sample and the beneficial mutation is introduced again until a successful sweep is observed. Note that our initial condition with each site polymorphic is not supposed to imply that the product of the population size and the per site mutation is much larger than one. Instead, those loci correspond to the subset of polymorphic loci scattered along a genomic region. The results do not depend on the number of segregating sites in the initial sample, as long as the pairwise difference between two randomly chosen genomes is almost always much larger than one. We repeated the simulations using only a tenth of the segregating sites and obtained similar results (see Supplement).

The Malthusian fitness of a haplotype is given by $F = s$ or $0$, depending on whether or not the haplotype carries the beneficial allele at position $L/2$. In each generation, a pool of gametes is produced to which each individual contributes a Poisson distributed number of copies with mean $\exp(F-\langle F \rangle + C)$, where $C = (1-N/N_0)$ serves as a carrying capacity to keep the population size approximately constant at $N_0$ and $\langle F \rangle$ is the mean fitness in the population. A fraction $r$ of the newly produced gametes are paired at random and crossed over at a randomly chosen position (since $r\ll 1$, there is no crossover in most gametes). At the $L-1$ neutral loci, mutations are introduced at random positions into the gametes with the specified mutation rate $u$. When simulating the sweeps of different strength, the mutation rate is changed accordingly to keep the ratio of the strength of selection and the total mutation rate in the range specified in the figure. This is akin to changing the size of the locus simulated, and allows more efficient simulations. The code is available on request.

To compare our method to previously developed methods to estimate sweeps \citep{Nielsen:2005p42523}, we generated data using the program \texttt{msms} by \citeauthor{Ewing:2010p43021}, see supplementary material.

\subsection*{Analysis of HIV data}

Sequences from the relevant samples were read by a python script and translated.  Sequences corresponding to identical amino acid sequences were grouped together. Within each of these groups, the number of transitions and transversion that do not change the amino-acid sequence were determined. In case the sequence overlapped with the end of the gene and extended into the long terminal repeat (LTR), only the aminoacid sequence up to the stop codon was used to group sequences and mutations in the LTR treated as neutral mutations. Sequences from the study on early immune system escape mutations by \citet{Fischer:2010p40314} are available from the NCBI short read archive under accession number SRA020793.  Sequences from the study on drug resistance evolution by \citet{Hedskog:2010p36144} were obtained from the authors.

\section*{Results}

\subsection*{Haplotype frequency spectrum of a selective sweep}
Consider a new adaptive mutation with selection coefficient $s>0$ in a panmictic haploid population of constant size $N$. We assume that selection is strong ($Ns\gg 1$). Once an adaptive mutation becomes established in the population by reaching a population frequency $x\approx (2Ns)^{-1}$ that assures its escape from future stochastic loss~\citep{Maynardsmith:1971p38196}, its frequency trajectory can be modeled by logistic growth
\begin{equation}
\label{eq:allelefrequency}
x(t)=\frac{e^{st}}{e^{st}+2Ns}.
\end{equation}

Let us first assume that recombination is infrequent compared to mutation (we will discuss recombination below). If neutral mutations occur during the early phase of the sweep on the adaptive haplotype, they can generate new variants of the adaptive haplotype that can also rise in frequency (\FIG{sweep_illustration}A). We adopt an infinite haplotypes model where every such mutation creates a distinct haplotype. When the frequency $x(t)$ of the beneficial allele is still small, novel adaptive haplotypes become established in the population at an approximate rate $\alpha(t)= 2s\times\mut Nx(t)$, where $\mut$ is the rate at which neutral mutations occur on the sweeping haplotype per generation. The factor $2s$ accounts for the establishment probability of those new haplotypes. After the sweep is completed, a novel haplotype variant will be the more frequent the earlier it arose in the sweep. Thus, to understand the spectrum of haplotype frequencies, we have to study the distribution of times at which these haplotypes are seeded.

Since novel haplotypes are seeded in independent events, the total number of established new haplotypes up to time $t$ is Poisson distributed with parameter
\begin{equation}
\label{eq:lambda}
\lambda(t) =\int_0^{t} \alpha(t')dt'
\approx \frac{\mut}{s}e^{st}.
\end{equation}
The approximation assumes that the relevant haplotypes are seeded while the sweeping allele is still rare and increases exponentially, which is appropriate as long as $t\ll \log(Ns)/s$, and that $s\gg u$, which can be ensured by reducing the size of the locus under investigation.

The probability density that haplotype $i$ is seeded at time $t$ is given by the rate $\alpha(t)$ of establishing new haplotypes multiplied by the probability of having $i-1$ haplotypes at time~$t$
\begin{equation}
\label{eq:dis_of_t}
p_i(t) = \alpha(t) \frac{e^{-\lambda(t)}\lambda(t)^{i-1}}{(i-1)!}
\propto e^{-\frac{\mut}{s}e^{st}+ist},
\end{equation}
where we have used the same approximations as in \EQ{lambda} and dropped factors independent of $t$ that ensure normalization. This distribution $p_i(t)$ is shown in \FIG{sweep_illustration}B for different $i$. The intervals between the peaks of $p_i(t)$ and $p_{i+1}(t)$ become smaller with increasing $i$, while the width of each peak decreases. More precisely, we derive from \EQ{dis_of_t} that the most likely seeding time of the $i$th haplotype is given by
\begin{equation}
t_i^* = \frac{1}{s}\log\left(\frac{is}{\mut}\right),
\label{eq:origtime}
\end{equation}
and that the peak of $p_i(t)$ has a width of $\sigma_i \approx (s\sqrt{i})^{-1}$ (if $i\gg 1$). The joint distribution of seeding times as well as the unordered frequency spectrum are derived in Supplementary Information, section 2.

Assuming that the $i$th haplotype establishes at $t_i^*$ and has same fitness as the founding haplotype, it will also rise in frequency, albeit with a time-lag $t_i^*$. Together with \EQ{origtime} we obtain an approximation for the expected frequency of the $i$th haplotype
\begin{equation}
\label{eq:hapfrequency}
x_i(t)=\frac{e^{(s-\mut)(t-t_i^*)}}{e^{s t}+2Ns} \to  e^{-\mut t}\left(\frac{\mut}{is}\right)^{1-\frac{\mut}{s}} \ .
\end{equation}
The limit corresponds to the asymptotic behavior as the beneficial allele approaches fixation.
Note that this expression holds for haplotypes $i\ge1$. The zero-th haplotype, i.e.,~the haplotype the mutation inititally arose on, grows as $x_0(t)=(e^{(s-\mut)t})/(e^{s t}+2Ns)$ and asymptotes to $e^{-\mut t}$. Each of the haplotype variants carrying the adaptive allele thus grows at a slightly smaller rate, $s-\mut$, than the overall rate $s$ at which the adaptive allele rises, accounting for the founding of new haplotype variants by mutation.

The interplay between the exponential amplification of the adaptive allele and the generation of new adaptive haplotype-variants thus gives rise to a simple power-law decay of the rank-frequency spectrum $\{x_0,x_1,\cdots\}$ of adaptive haplotypes: the most abundant adaptive haplotype, on average, is typically $s/\mut$ times more frequent than the 2nd most abundant adaptive haplotype, $2s/\mut$ times more frequent than the 3rd most abundant adaptive haplotype, and so forth. This power-law spectrum with exponent $\beta =1-\mut/s$ differs markedly from that of neutral evolution, where haplotype frequencies are expected to decay as $x_i/x_0\sim e^{-i/\Theta}$ with haplotype rank $i$ ($\Theta = 2Nu$; see Supplementary Information, section 1).

The distribution of haplotype frequencies after a sweep is related to the distribution of ``family sizes'' studied by \citet{Barton:1998p28270}. \citeauthor{Barton:1998p28270} presented numeric results for the distribution of the size of groups of individuals that share a common ancestor at the beginning of the sweep. Haplotype frequencies, however, refer to the size of groups identical by state, rather than descent, and characterize diversification that happened after the sweep, rather than the ancestral variation that survived the sweep.

\begin{figure}[h!]
\begin{center}
\includegraphics[width=0.49\columnwidth,type=pdf,ext=.pdf,read=.pdf]{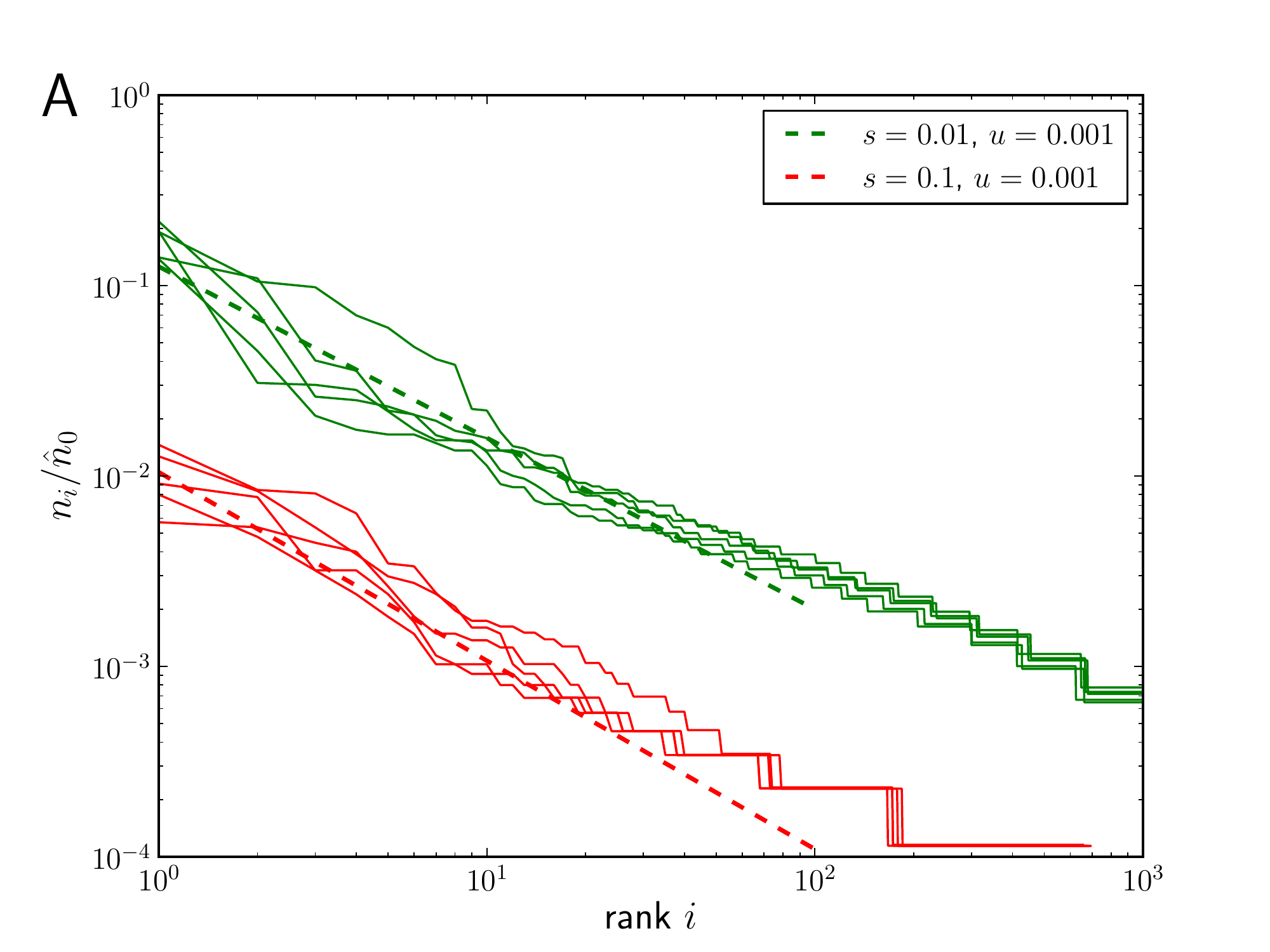}
\includegraphics[width=0.49\columnwidth,type=pdf,ext=.pdf,read=.pdf]{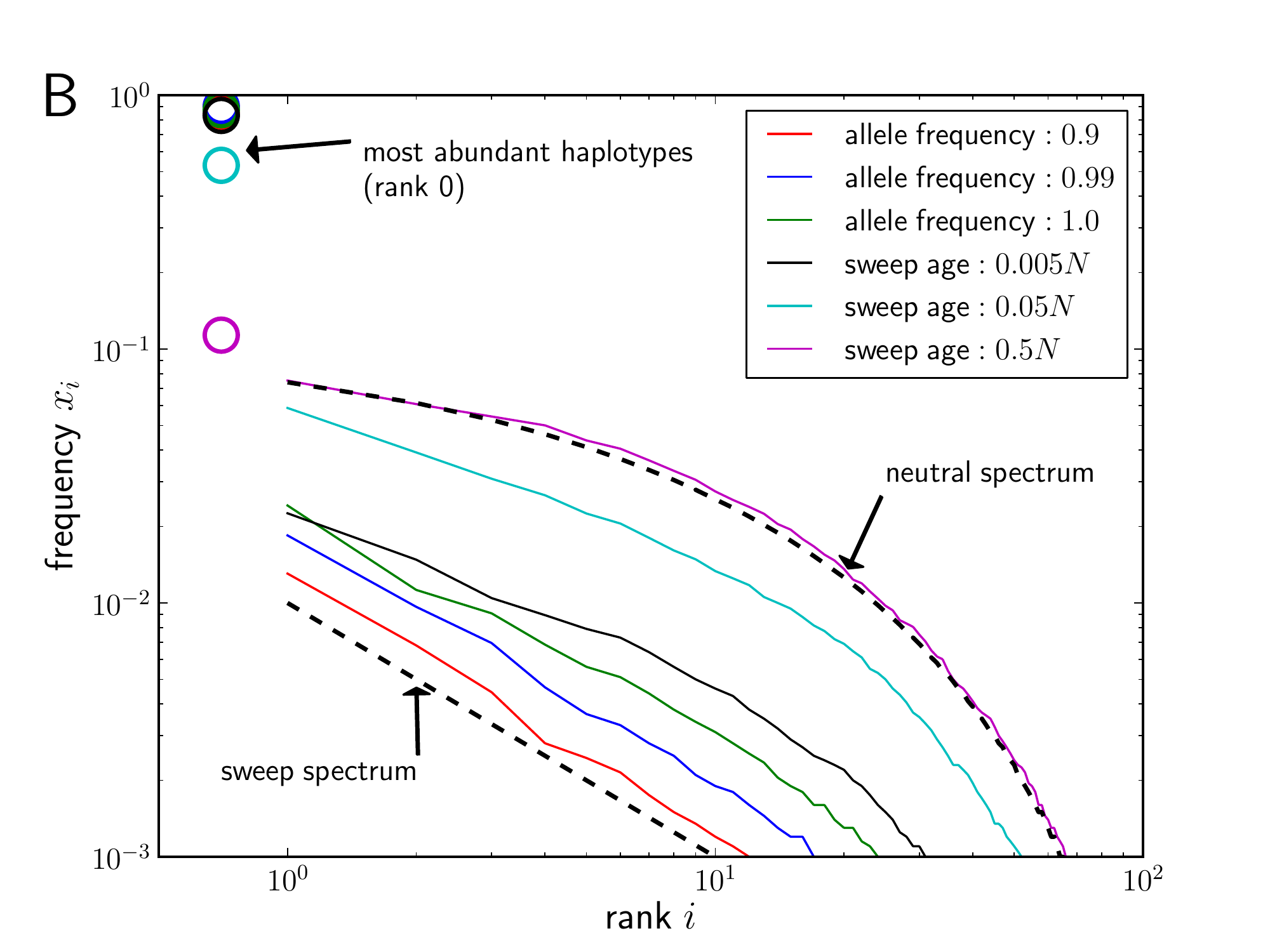}
\caption[labelInTOC]{
(A) Ensembles of $5$ haplotype rank-frequency spectra for $\mut=0.001$ and two selection coefficients $s=0.1$ and $s=0.01$. Spectra were estimated from samples of size $n=10^4$ taken when the allele frequency of the beneficial mutation reached $0.99$. The deterministic expectations given in \EQ{rank} are shown as dashed lines. The normalizer $\hat{n}_0$ was calculated according to \EQ{nhat} with $k=4$.
(B) Relaxation of haplotype spectra to the neutral spectrum. Rank-frequency spectra were estimated from samples of size $n=10^4$ when the beneficial allele reached population frequencies 0.9, 0.99, 1.0, and for several time points after completion of the sweep (time in generations) ($s=0.001$, $\mut=10^{-5}$, $L=1024$, $N=10^6$, $\Theta=2Nu=20$). Each curve is the median of 50 simulated sweeps. The most common haplotypes (rank 0, not represented on log scale) are indicated by circles. Right after the sweep, the population consists of one very common haplotype, $x_0\approx 0.98$, and a large number of rare variants whose frequency decays as predicted by \EQ{rank} (lower dashed line). Over time of order $t\approx N$, the frequency spectrum relaxes to the expected neutral spectrum (upper dashed line). }
\label{fig:sweep_spectra_relaxation}
\end{center}
\end{figure}

In order to compare the haplotype counts $n_0>n_1>n_2\ldots$ in a sample of size $n$ to \EQ{hapfrequency}, we need an estimate of $e^{-\mut t}$. The simplest estimate of $e^{-\mut t}$ is the sample frequency of the most abundant haplotype, $n_0/n$, whose deterministic value would be precisely $e^{-\mut t}$. However, $n_0$ can be quite variable because the variances of the first few seeding times are still large. To circumvent this problem, one can construct a more stable estimate of $e^{-\mut t}$ based on the first $k$ haplotypes and compare the sum  of the sample frequencies to the sum of the deterministic expectation of the frequencies:
\begin{equation}
\label{eq:nhat}
\hat{n}_0 = \frac{\sum_{i=0}^k n_i}{1+\sum_{i=1}^{k}\left(\frac{\mut}{s i}\right)^\beta} \ ,
\end{equation}
where $\beta = 1-u/s$. This effective $\hat{n}_0$ is insensitive to $k$ as long as the first few haplotypes are included. \FIG{sweep_spectra_relaxation}A shows a comparison of the simulated haplotype spectra with
\begin{equation}
\label{eq:rank}
\frac{n_i}{\hat{n}_0}\approx \left(\frac{\mut}{is}\right)^\beta \ .
\end{equation}
The logarithmic axes of the plot are chosen such that the algebraic decay shows up as a slope $-\beta$. The observed slope is correct initially, but the curves become slightly flatter after haplotype rank $\approx 50$. This is due to degradation of the sweep pattern by random genetic drift, which we will discuss next.

In deriving \EQ{rank}, we have neglected fluctuations in the seeding and establishment times of novel haplotype variants as well as genetic drift after the sweep is complete. Relaxing these assumptions requires a stochastic calculation, which is given in the Supplementary Information, sections 3A and B. Specifically, we calculate the probability of finding $\is$ haplotypes to be present in more than $\ns$ copies in the population, assuming the logistic trajectory of the beneficial allele given in \EQ{allelefrequency}. These calculations are lengthy, but the resulting effects can be understood by the simple arguments given below.

Variations in the establishment times will cause the haplotype spectra to fluctuate and result in spectra below or above \EQ{rank} if new haplotypes were seeded fortuitously early or late. As expected from the distribution of seeding times in \FIG{sweep_illustration}, the frequencies of the first few haplotypes fluctuate substantially, while the later ones do not. Since the number of haplotypes seeded by time $t$ is Poisson distributed, so is the number of haplotypes above a certain frequency after the sweep is complete (neglecting genetic drift, there is a one to one mapping between establishment time and eventual frequency). Hence, the fluctuations due to random seeding times are small if the expected number of haplotypes above the chosen frequency is large. In accordance with these arguments, variation of haplotype spectra in \FIG{sweep_spectra_relaxation}A decreases with increasing $i$.

Once the beneficial allele approaches fixation, the strength of selection goes to zero, and exponential amplification ceases. The frequencies of the rare haplotypes thereafter change primarily due to random drift, while the frequencies of the common haplotypes decrease due to additional mutations that produce new rare haplotypes. Both of these processes lead to a homogenization of haplotype frequencies, i.e., common haplotypes becoming rarer and rare haplotypes becoming more common, which ultimately wipes out the sweep signature. This relaxation to the neutral haplotype spectrum is shown in \FIG{sweep_spectra_relaxation}B.

Since the time required to drift to frequency $\xs$ is approximately $N\xs$, the haplotype rank-frequency spectrum will soon develop a bulge of drift dominated haplotypes at low frequency and high rank. This accumulation of rare haplotypes due to genetic drift can be calculated perturbatively (see Supplementary Information, section 3). For the expected number of haplotypes $\is$ with frequencies above $\xs$, one obtains
\begin{equation}
\label{eq:hapspectra_with_drift}
\langle \is\rangle \approx
\frac{\mut}{s}\left[\frac{e^{-\mut t}}{(\xs-\Delta t/N)}\right]^{1+\mut/s} \ ,
\end{equation}
where $\Delta t$ is the age of haplotype $\is$ which is of the same order as the age $t$ of the sweep.
Note that \EQ{hapspectra_with_drift} is essentially \EQ{hapfrequency} solved for $i$ with the ``drift contribution'' $\Delta t/N$ subtracted from the frequency $\xs$.
Since the age $\Delta t$ of the haplotype is similar to that of the beneficial allele, $t$,  we conclude that for frequencies less than $t/N$, the sweep spectrum is eroded, while for frequencies much larger that $t/N$, it prevails. After a time of order $N$, the entire spectrum has relaxed to the neutral haplotype frequency spectrum as shown in \FIG{sweep_spectra_relaxation}B. Note that the time it takes for the beneficial allele to go from frequency 0.9 to complete fixation can be long, and substantial erosion of the sweep spectrum can happen during this time interval.

In general, the time that has elapsed since the beginning of the sweep is unknown. To obtain an estimate of the age of the sweep, it is useful to reconsider \EQ{hapfrequency}, which states that the frequency of the founding haplotype is expected to asymptote to $e^{-\mut t}$. This behavior, and in particular the more accurate estimate of $e^{-\mut t}$ given in \EQ{nhat}, allows one to obtain a rough estimate of the sweep's age.

In addition to genetic drift, limited sampling of the population results in a deviation of the observed from the expected sweep spectra. The detected haplotype counts are a convolution of the their true frequencies with the distribution expected from sampling the population. After ranking of haplotype counts, a large number of rare variants leads to a flattening of the spectrum, and in particular an excess of singletons.

\subsection*{Estimating selection strength}

According to \EQ{rank}, the expected rank-frequency spectrum of haplotypes in a selective sweep is determined by the single parameter $\mut/s$ -- the ratio of the mutation rate over the locus and the strength of selection. Given an independent estimate of $\mut$, one can therefore use the haplotype frequency spectrum to estimate the selection coefficient of a sweep, for example by simply counting the number of different haplotypes present above a specified frequency cutoff in the sample. Let $\is$ be the overall number of different haplotypes which are present in at least $\ns$ copies each. The estimator for the strength of selection is then
\begin{equation}
\hat{s} = \frac{\mut}{\is} \left(\frac{\hat{n}_0}{\ns}\right)^{1+\frac{\ns\is}{\hat{n}_0}} \ ,
\label{eq:estimator_s}
\end{equation}
where $\hat{n}_0$ is either set to the observed $n_0$ or determined via \EQ{nhat}.

The estimator $\hat{s}$ allows us to either fix $\ns$ and then determine the count $\is$ from the data, or fix $\is$ and then measure $\ns$. In either case, the cut-off should be chosen to achieve maximum accuracy. Since the number $\is$ of haplotypes above a frequency threshold is Poisson distributed, fluctuations of $\hat{s}$ decrease with smaller $\ns$ and larger $\is$. To low $\is$, however, will include parts of the frequency spectrum that has already been degraded by drift, which predominantly affects the rare haplotypes. In addition, limited sampling depth limits $\is$. Hence $\is$ should be chosen as large as possible, but such that $\ns\gg 1$ and the frequency spectrum is still of power-law form down to haplotype $\is$. In this case, the relative error of $\hat{s}$ is of order $1/\sqrt{\is}$ (assuming the model assumptions are met).

\begin{figure}[h!]
\begin{center}
\includegraphics[width=0.49\columnwidth,type=pdf,ext=.pdf,read=.pdf]{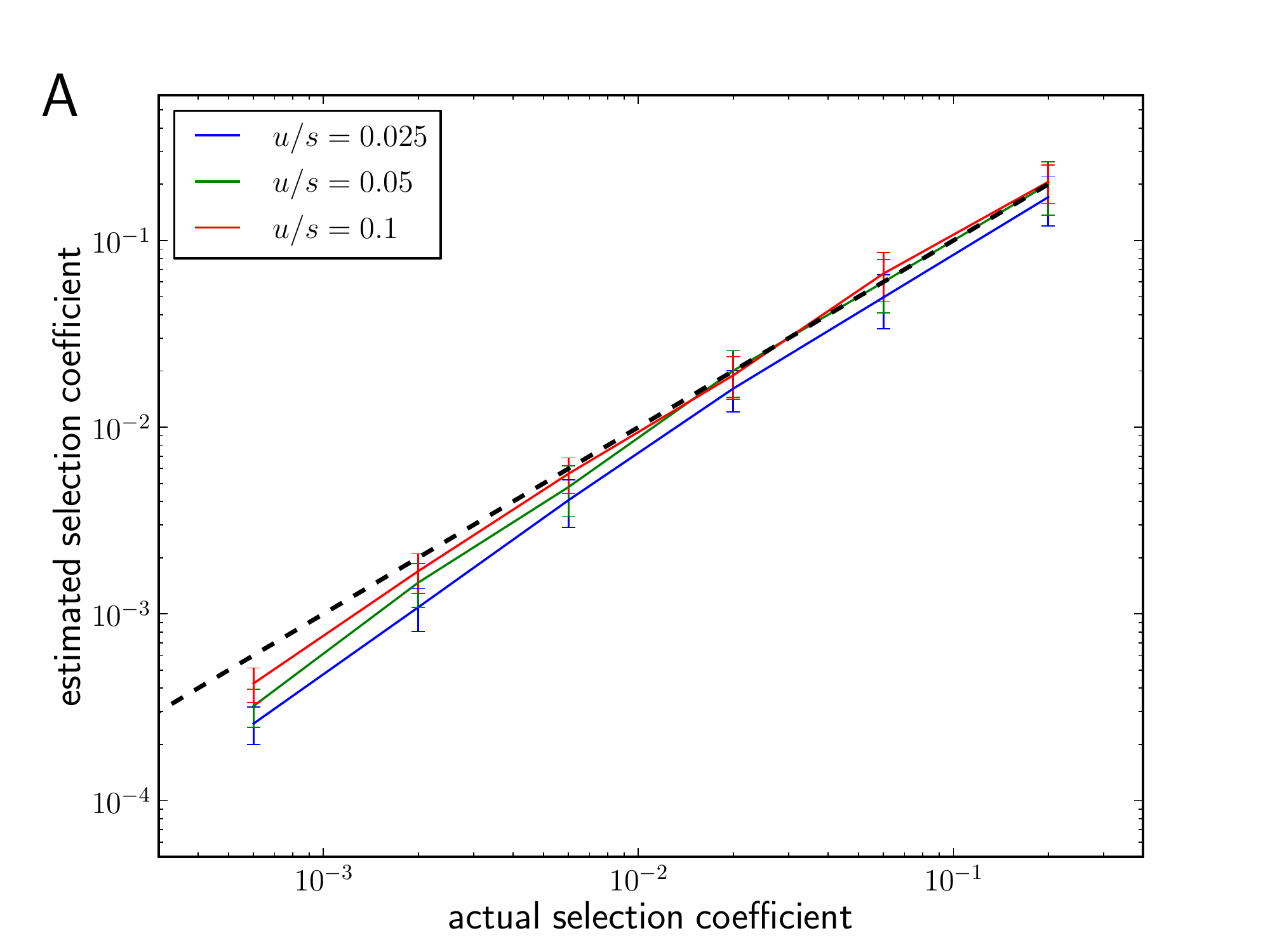}
\includegraphics[width=0.49\columnwidth,type=pdf,ext=.pdf,read=.pdf]{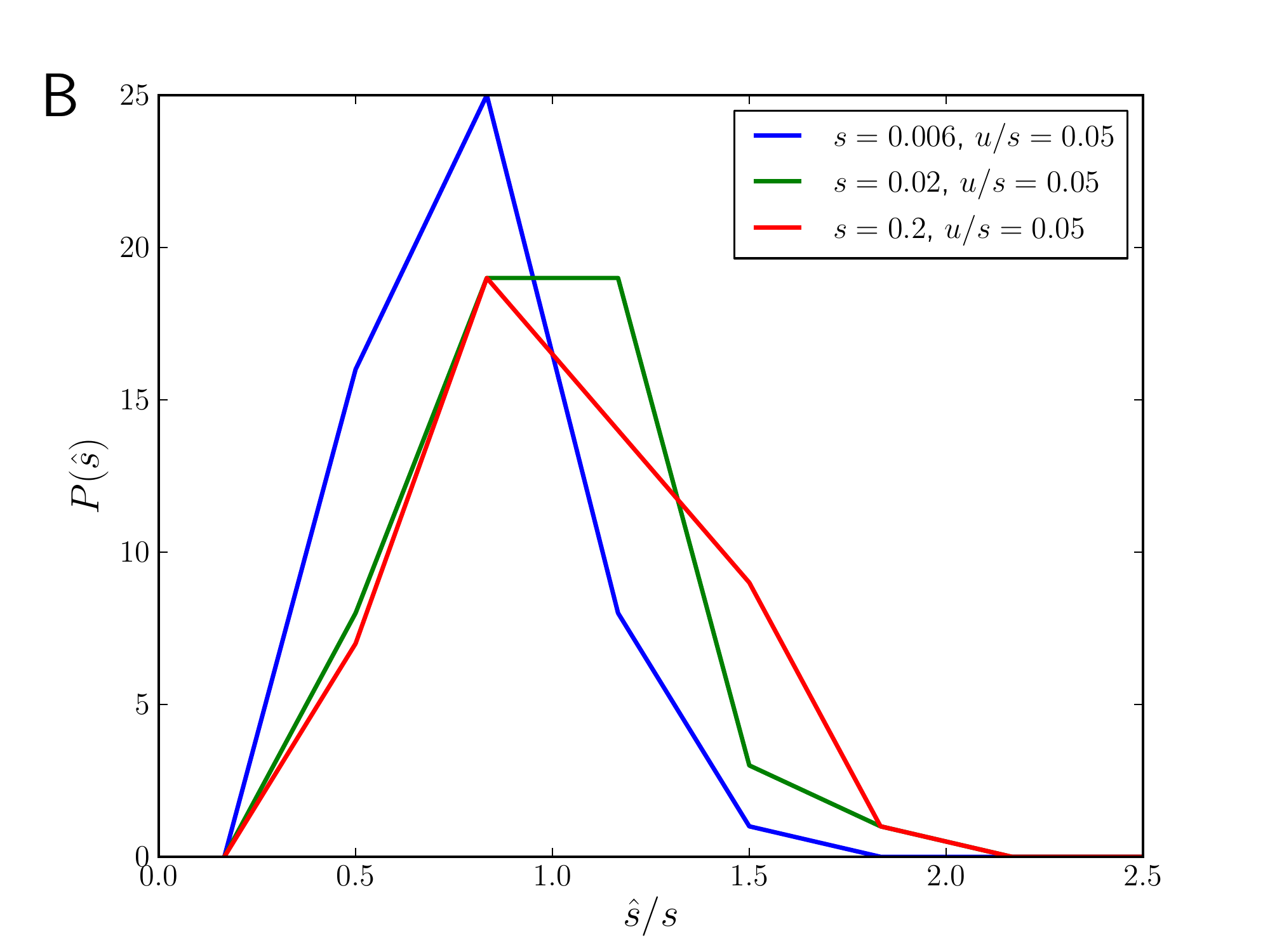}
\includegraphics[width=0.49\columnwidth,type=pdf,ext=.pdf,read=.pdf]{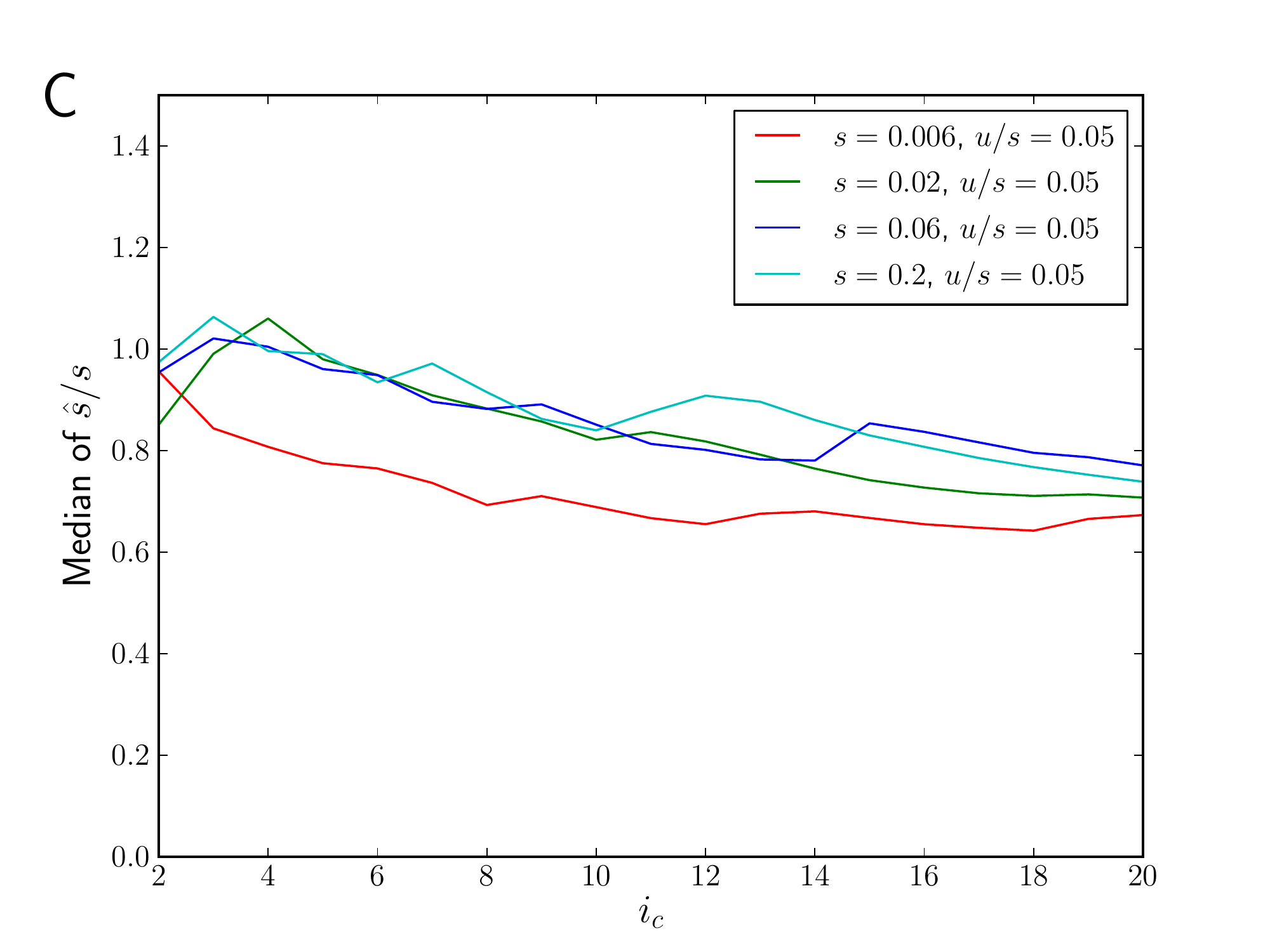}
\includegraphics[width=0.49\columnwidth,type=pdf,ext=.pdf,read=.pdf]{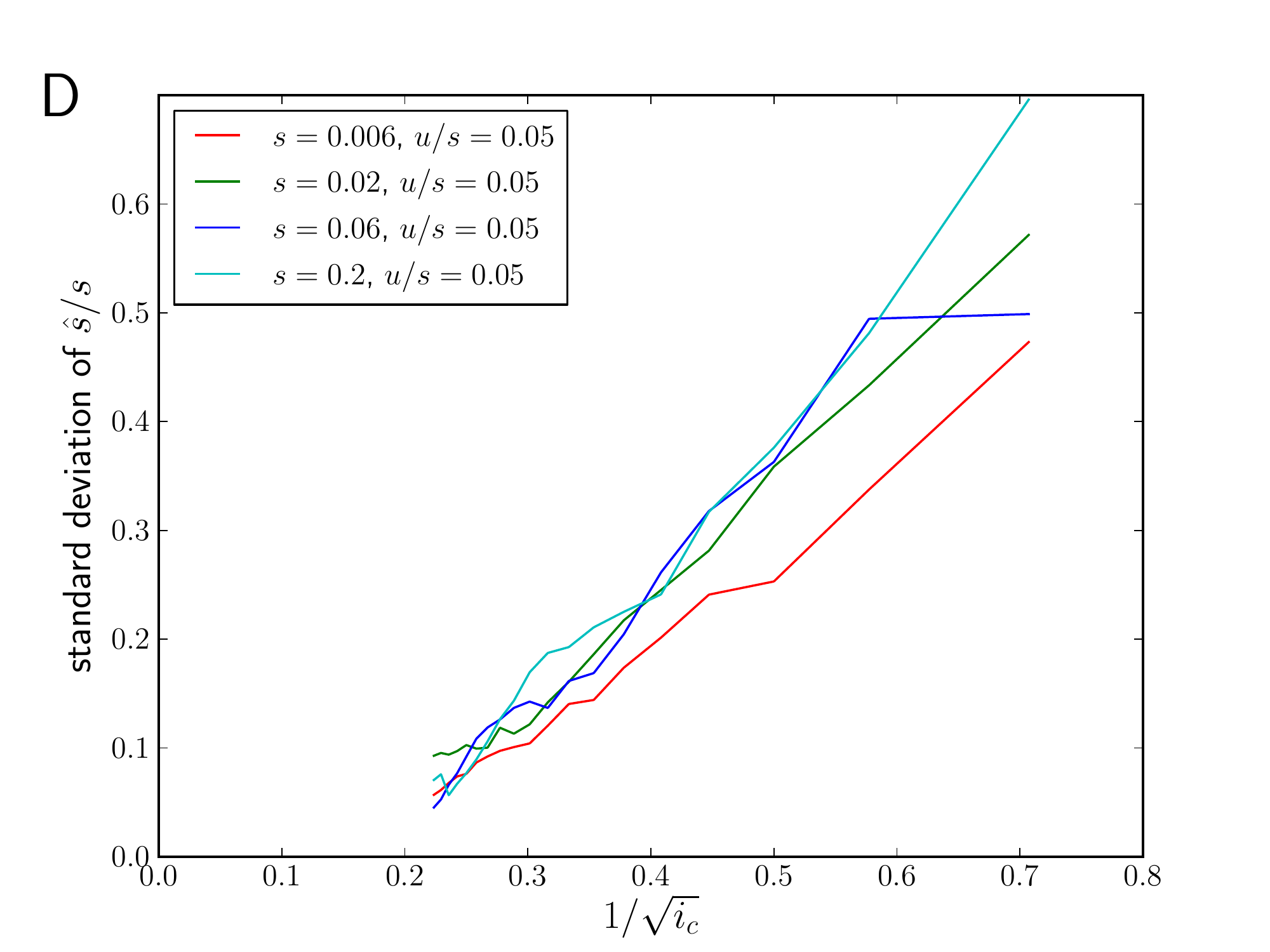}
\caption[labelInTOC]{(A) Mean $\hat{s}$ over 50 simulation runs for a wide range of selection strengths and three different ratios of $\mut/s$ using \EQ{estimator_s} with $\is=5$. Error bars are $\pm$ one standard deviation. The dashed black line indicates the correct $s$. Population samples of size $n=10^3$ were taken when the beneficial allele reached fixation. For small $s$, there is a slight downward bias due to genetic drift. $L=4096$ and $N=10^6$.  (B) Distribution of $\hat{s}$ at different selection strengths for fixed $\mut/s=0.05$. (C) Performance of $\hat{s}$ as a function of $\is$. There is a tendency to underestimate the strength of the sweep if $\is$ is large, which arises from the inclusion of rare haplotypes that are affected by genetic drift and may contain haplotypes that arose after the sweep. (D) Standard deviation over $50$ estimates as a function of $\is^{-1/2}$. Variation is suppressed at large $\is$ due the discreteness of low population frequencies.}
\label{fig:sweep_relaxation_estimation}
\end{center}
\end{figure}

It is generally advisable to fix $\is$ and determine $\ns=n_{\is}$ because spectra come flatter than $i^{-1}$ due to the confounding effect of genetic drift and limited sampling of the population. \FIG{sweep_relaxation_estimation} shows the performance of our estimator when applied to simulated sweeps (see Methods) for different selection coefficients and mutation rates as well as its dependence on the choice of $\is$. The simulations confirm that our estimator performs accurately over the range of moderate selection ($s\approx 10^{-3}$) to strong selection ($s\approx 0.2$) given the parameters used. However, there is a systematic downward bias for small $s$. This bias is due to genetic drift and limited sampling depth.

In the Supplementary Information, section 3, we derive an estimator that accounts for genetic drift perturbatively, see Equation (24) of the supplementary information.  In essence, this estimator contains the same correction present in \EQ{hapspectra_with_drift}, i.e., it subtracts the drift contribution $t/N$ from the frequency of the rare haplotypes. Importantly, this drift correction sets the limits of applicability of $\hat{s}$: (i) After the completion of the sweep, we can estimate $s$ only for a limited time. Specifically, for a given sample frequency $\ns/n$, we require that $t < N\ns/n$. (ii) The range of $s$ that can be reliably estimated by the method is limited: even if we catch the sweep right at the time of fixation of the adaptive allele, it still needs to have occurred faster than the time scale of its own degradation by drift. Given that the time it takes to complete a sweep is approximately $(2/s)\log Ns$ and that $\ns/n\approx \mut/(s\is)$, we require $N> (2\is/u)\log Ns$. Together with $s\gg u$, we find that the range of $s$ that can be estimated is bounded from below by $N s \gg \is \log Ns$. This estimate is consistent with the deviations in \FIG{sweep_relaxation_estimation}  for $s<10^{-3}$.

\subsection*{Recombination}

Up to now, we have neglected any potential effects of recombination on the frequency distribution of the new adaptive haplotype variants arising in a selective sweep. In fact, it is one of the advantages of our estimator that it does not exclusively rely on recombination for its inference of selection coefficients and can thus also be applied to systems which lack recombination or where recombination rates are not well known. However, if recombination occurs on the sweeping haplotype, this can generate new variants of the adaptive haplotype in a manner similar to mutation. This becomes important whenever the recombination rate $r$ over the locus is comparable to its mutation rate $\mut$. In this case, the rate at which new haplotype variants are generated during the sweep will actually be higher than assumed based on the mutation rate $\mut$, and we thus expect $\hat{s}$ to underestimate the strength of the sweep.  

\begin{figure}[htp]
\begin{center}
\includegraphics[width=0.49\columnwidth,type=pdf,ext=.pdf,read=.pdf]{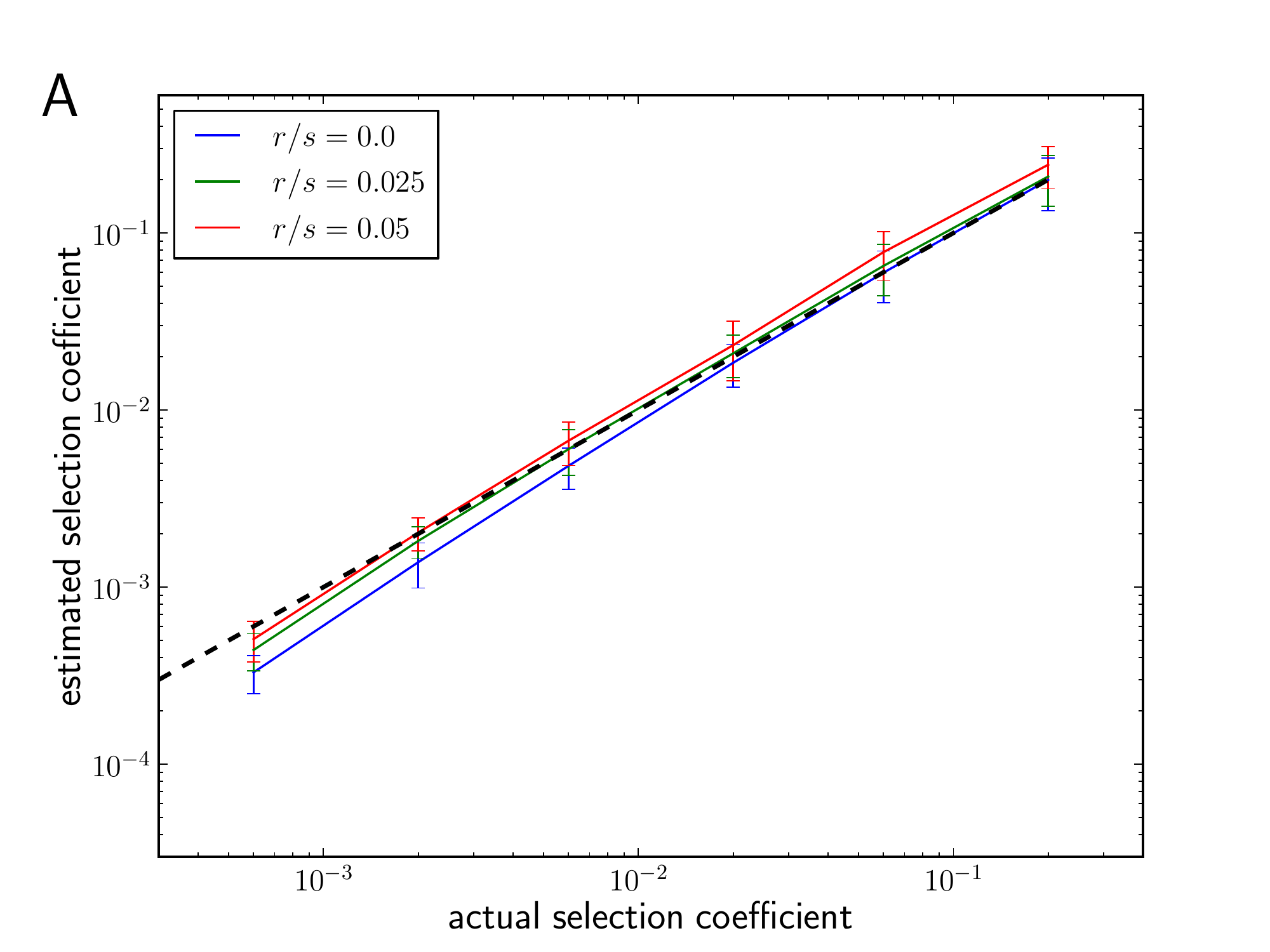}
\includegraphics[width=0.49\columnwidth,type=pdf,ext=.pdf,read=.pdf]{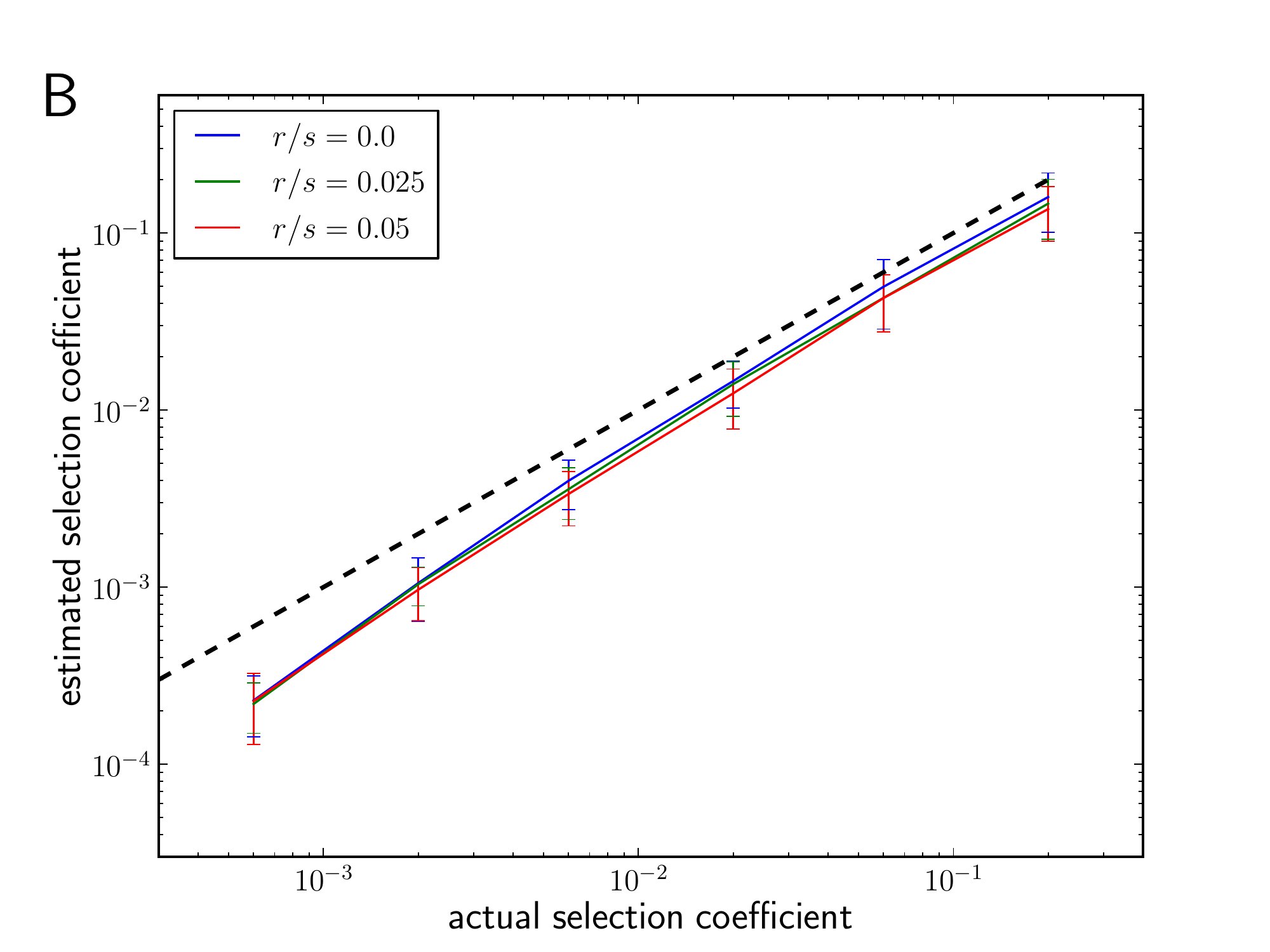}
\caption[labelInTOC]{Estimating selection coefficients in the presence of recombination. (A) Performance of our estimator $\hat{s}$ using \EQ{estimator_s} with diversification rate $\mut+r$ for different $r$ with $\mut/s=0.05$. The slight overestimation at high recombination rates is due to fact that high and low frequency haplotypes degrade at slightly different rates due to the possibility of recombination with an identical haplotype. (B) Estimation of $\hat{s}$ from haplotype spectra restricted to those haplotypes that differ at only one site from the most abundant haplotype. After pruning, there are fewer abundant haplotypes and the estimates are more sensitive to genetic drift.}
\label{fig:estimator_rec}
\end{center}
\end{figure}

We propose two ways to incorporate recombination: First, one can treat recombination analogously to mutation and assume that every recombination event generates a new variant of the adaptive haplotype that will be different from all other existing variants. Under this assumption, the mutation rate $\mut$ simply needs to be replaced by the sum $\mut+r$. \FIG{estimator_rec}A shows how our estimator performs for different ratios of mutation to recombination rate under this approach. The infinite haplotypes assumption is appropriate if ancestral diversity is high and independent recombination events are thus unlikely to yield equal outcomes. If, however, ancestral diversity is low, the infinite haplotypes model will overestimate the rate at which new haplotypes arise.

Alternatively, we can try to effectively exclude most recombined adaptive haplotype variants. Recombined variants are likely to differ from the founding haplotype at more than one site, whereas variants originating from mutation should differ at only one site. By filtering out the former, we can treat the remainder as originating primarily from mutation events. As shown in the Supplementary Information, section 4, such a set of pruned ``mutation-only'' haplotypes has a slightly different rank-frequency spectrum: The exponent $\beta$ is exactly 1, rather than $1-\mut/s$. Besides that, \EQ{estimator_s} can be applied with $\mut$ being the mutation rate and $\ns$ being the abundance of the $\is$ haplotype in the pruned set of haplotypes. The estimates obtained by this procedure are shown in \FIG{estimator_rec}B. While this modified estimator still seems to underestimate selection coefficients slightly, it performs consistently across different recombination rates.

The difference in the exponent of the rank frequency spectrum arises from slightly different rates of seeding and amplification of new haplotypes. Without pruning, the rate of establishment of novel haplotypes is proportional to the frequency of the beneficial allele, which follows a logistic growth with rate $s$. Haplotype frequencies, however, only grow with rate $\sur = s - \mut -r$. This slight difference in rates is responsible for the deviation of the exponent $\beta$ from $1$ by $\mut/s$ or $(\mut+r)/s$, where the latter also accounts for recombination. When restricting the analysis to haplotypes that descend directly from the founding haplotype, the rate of establishment of novel haplotypes is proportional to the frequency of the founding haplotype. Hence, establishment and amplification happen with the same rate $\sur = s - \mut -r$ and the exponent is exactly 1.

We note another minor difference in the dynamics of haplotype frequencies in the presence of recombination. When the beneficial allele reaches fixation, the most abundant haplotype will most likely recombine with itself. Thus it diversifies more slowly after the sweep is complete. Minor haplotypes, however, continue to diversify through recombination, which has the effect of slowly decreasing their frequency and producing novel low frequency haplotypes. This slow decrease opposes the effects of genetic drift, which has the tendency to increase the frequency of minor alleles. In simulations, estimates of $s$ often become more accurate due to this effect.

\subsection*{Application of $\hat{s}$ to deep sequencing data from HIV}

Deep population diversity data of the kind required for the application of our estimator have recently been obtained for HIV populations  by sequencing viral RNA from plasma samples \citep{Tsibris:2009p26554,Fischer:2010p40314,Hedskog:2010p36144}. We will present three examples from such studies to validate our method and discuss its applicability.

\begin{figure}[t!]
\begin{center}
\includegraphics[width=0.495\columnwidth,type=pdf,ext=.pdf,read=.pdf]{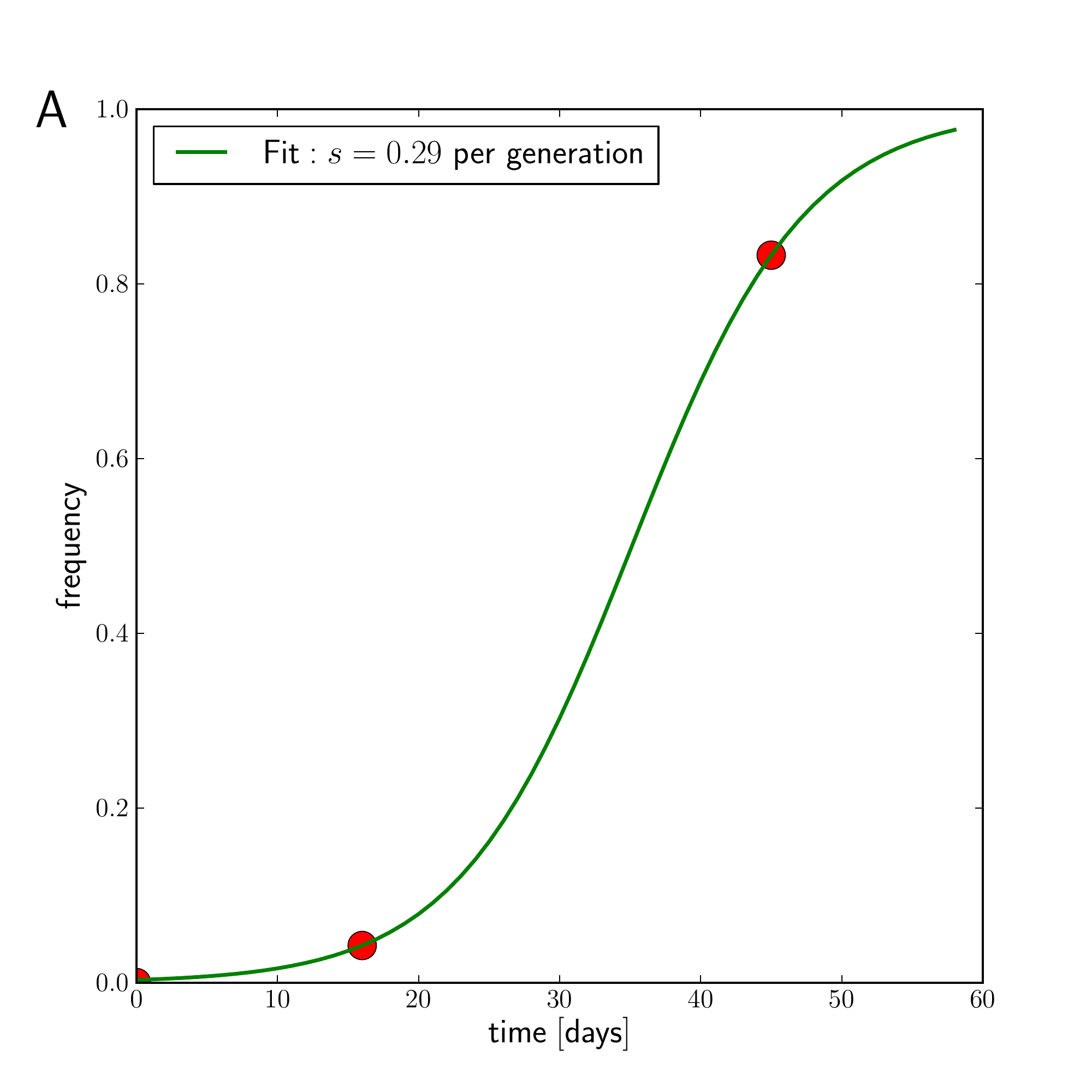}
\includegraphics[width=0.495\columnwidth,type=pdf,ext=.pdf,read=.pdf]{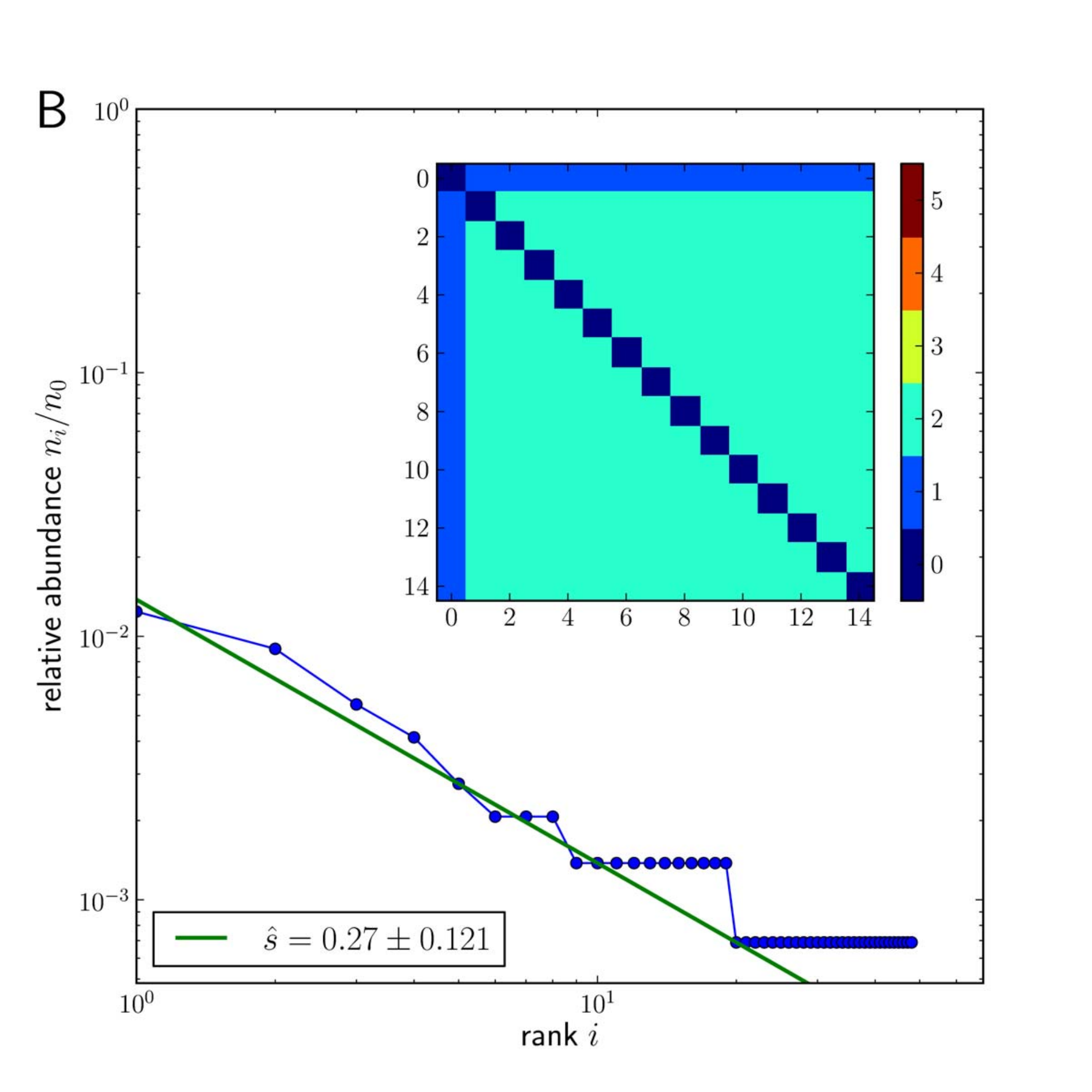}
\caption[labelInTOC]{The rise of a mutation selected by the immune system in the gene \gene{Nef} of patient CH40 from \citet{Fischer:2010p40314}. (A) Frequency trajectory of the adaptive allele with fitted curve for $s\approx 0.3$.   (B) Observed haplotype spectrum of variation due to synonymous transitions from all sequences containing the selected allele at day 45. The straight line indicates the $1/i$ behavior. The pairwise nucleotide distance between all haplotypes ordered by abundance is shown in the inset. The error of the estimate are obtained assuming Poisson statistics with $\is=5$. Note, however, that the uncertainty of the mutation rate is of the same order. }
\label{fig:HIV_example1}
\end{center}
\end{figure}

As a first example, we investigated a sample taken shortly after infection when HIV evolution is primarily driven by mutations that allow the virus to escape the immune system. For each of the patients studied in \citet{Fischer:2010p40314}, samples from several time points were investigated and amplicons sequenced using 454 technology. In one of the epitopes studied (gene \gene{nef} in patient CH40), a mutation spread sufficiently slowly such that it was possible to observe the mutation rise from low frequency at day 16 to high frequency at day 45 (\FIG{HIV_example1}A). From this time series \citet{Fischer:2010p40314} estimate a selection coefficient of 0.3, assuming a generation time of HIV between 1.5 and 2 days \citep{Perelson:1996p23158},

To validate our estimator, we can compare this direct estimate of the selection coefficient obtained from time-course data with that obtained from the frequency spectrum of haplotype diversity at the locus, which is shown in \FIG{HIV_example1}B. The figure contains the frequency of haplotypes with the dominant amino acid sequence which differ only by putatively neutral synonymous mutations. The haplotype rank-frequency spectrum according to \EQ{estimator_s} suggests a ratio of $u/s\approx 0.014$. There is considerable uncertainty in the mutation rate estimates for HIV, ranging from $3.4\times 10^{-5}$ \citep{Mansky:1995p38971} to $9\times 10^{-5}$ per site and generation (\citet{ONeil:2002p43468} for mutations in the LTR). The majority of these mutations are transitions. In our example the sequenced locus tolerates a total of 93 transitions that do not change the amino-acid sequence of \gene{Nef} or fall into the LTR which overlaps the sequenced region. Using $4\times 10^{-5}$ per generation as an averaged estimate for the transition rate, we arrive at
\begin{equation}
\hat{s} \approx \frac{93\times 4\times 10^{-5}}{0.014}\;\mathsf{gen}^{-1} = 0.27 \;\mathsf{gen}^{-1} \pm 0.12.
\end{equation}
The uncertainty stemming from the random times at which the haplotypes are seeded is expected to result in relative errors $\sim 1/\sqrt{\is}$, which for $\is=8$ used here amounts to $\pm 0.12$. Additional uncertainty (probably larger) in the mutation rate needs to be added to these error bars. Given these uncertainties we consider our estimate in excellent agreement with the independent estimate from time-course data. Note that recombination is not expected to make a large contribution to haplotype diversification since the effective HIV recombination rate is less than half the mutation rate~\citep{Neher:2010p32691,Batorsky:2011p40107}. Furthermore, the HIV population has undergone several sweeps prior to the date this sample was taken such that the ancestral diversity was very low.

\begin{figure}[t!]
\begin{center}
\includegraphics[width=0.495\columnwidth]{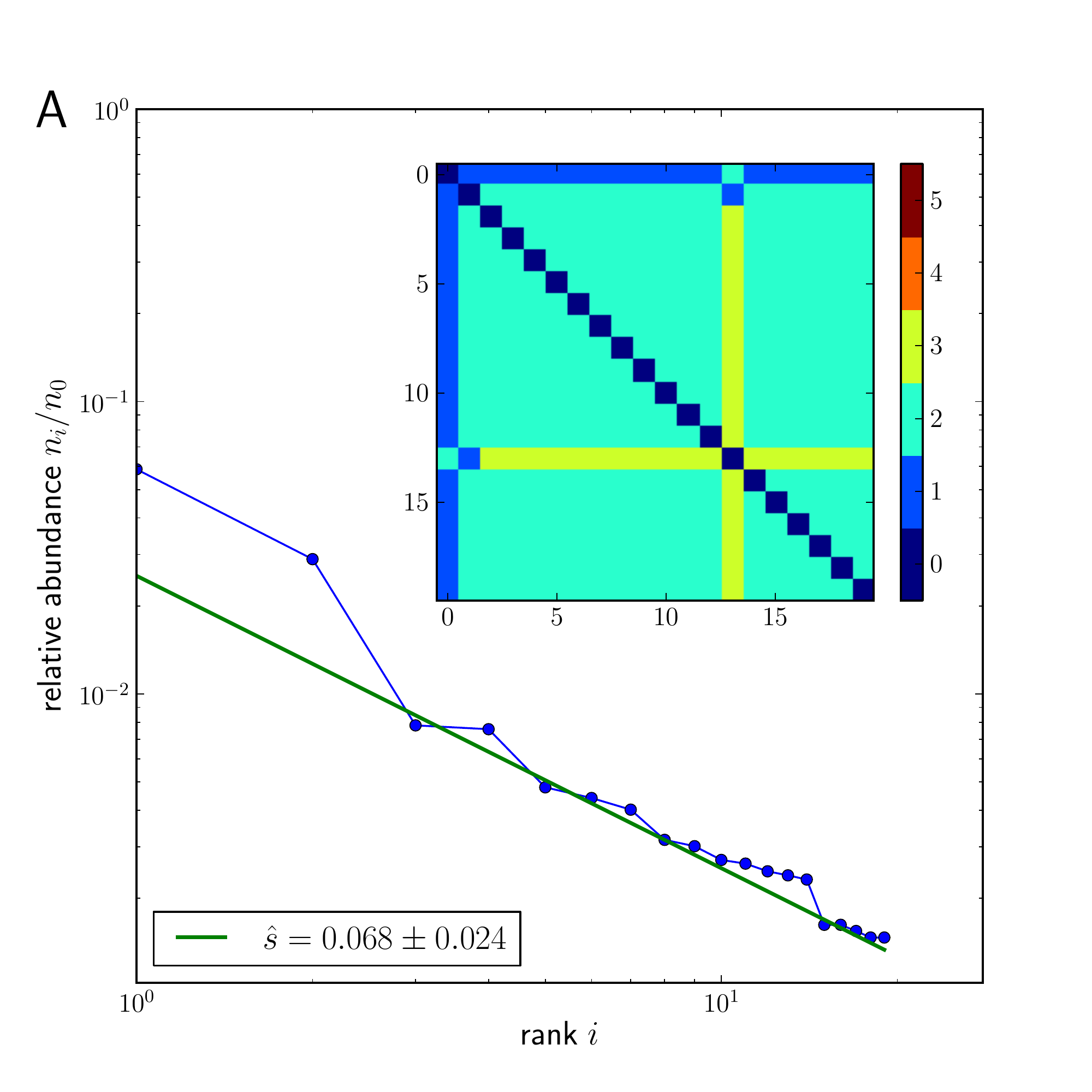}
\includegraphics[width=0.495\columnwidth]{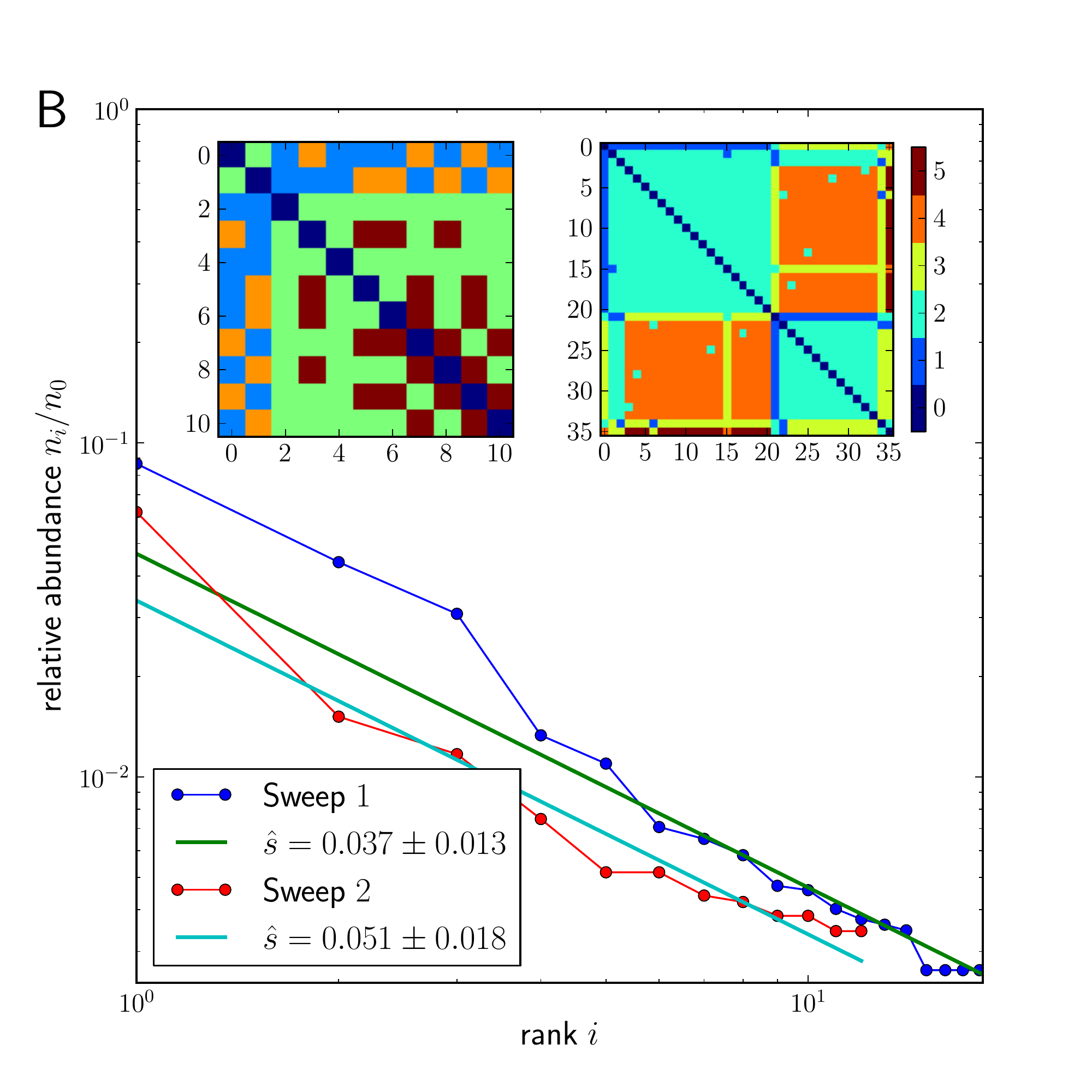}
\caption[labelInTOC]{(A) Exemplary sweep in HIV evolution. The rank-frequency spectrum and pairwise distance matrix are shown for the different haplotypes coding for the same amino-acid sequence aa 180-220 of the RT~\citep{Hedskog:2010p36144}, comp.~\FIG{HIV_example1}. All but haplotype 13 differ from the most abundant haplotype (0) by one mutation. Haplotype 13 is most likely a descendent of haplotype 1. (B) Example of a soft sweep in HIV evolution. In this sample, the two most abundant haplotypes (0) and (1) differ by more than one mutation (top left corner of left inset, showing the distance matrix of the 10 most abundant haplotypes), suggesting that the selected mutation arose independently on different backgrounds. By inspecting the first two columns of the left distance matrix, one finds that most of the other haplotypes differ by one mutation from either haplotype (0) or (1), consistent with the beneficial mutation having arisen independently on haplotypes (0) and (1). Assigning the rare haplotypes to either of the two abundant ``founding'' haplotypes and reordering yields the distance matrix shown as the right inset, this time including all 36 haplotypes. The reordered matrix displays two characteristic sweep patterns as blocks on the diagonal. Indeed, the haplotypes of each of these two blocks display the characteristic power-law rank-frequency spectrum. In both panels, $\is=8$ and relative errors are $1/\sqrt{\is}$.}
\label{fig:HIV_example2}
\end{center}
\end{figure}

We also checked whether the haplotype variation in our sample is consistent with the characteristic star-phylogeny expected for mutations that arose on the background of the sweeping founder haplotype. In particular, under the assumption of an infinite sites model where each such mutation gives rise to a new haplotype, we would expect that descendants should always be one mutation away from the founding haplotype, while any two descendants should be two mutations away from each other. Our example shows precisely this pattern, as can be seen in the inset of \FIG{HIV_example1}A. This is further evidence that recombination did not contribute substantially to haplotype diversification.

As a second example, we consider deep sequencing data of the reverse transcriptase (RT) locus of HIV (120 bp, amino acids 180-220) \citep{Hedskog:2010p36144}. The patients analyzed in this study were under anti-retroviral treatment, which implies strong selection pressure for resistance mutations at the RT locus. During the course of treatment resistance has evolved repeatedly in several patients. We therefore expect to see signatures of strong selective sweeps in these viral populations.

As before, we restricted our analysis to synonymous differences because our model assumes selectively neutral mutations. The 120 bp analysis windows typically contained enough variation such that in most samples more than 8 different haplotypes were sampled. Hence we choose $\is=8$ for estimating $\hat{s}$ and then measured the cutoff frequency $\ns$ of the least abundant haplotype to evaluate \EQ{estimator_s}.

\FIG{HIV_example2}A shows the measured haplotype rank-frequency spectrum for one of the patient samples (time point 4, patient 4). The spectrum decays with the characteristic power-law of a selective sweep. As discussed above, we determined the number of synonymous transitions in the founding sequence and used $4\times 10^{-5}$ as the rate of individual transitions. With this mutation rate estimate, we arrive at $\hat{s}\approx 0.07$ per generation. Note that the causative mutation, or the combination of mutations, that drove this sweep does not necessarily have to lie inside the analyzed window but could also be located elsewhere in the genome.

\FIG{HIV_example2}B shows haplotype data from another population sample (time point 4, patient 5). In this third example, the pairwise mutational distances between different haplotypes reveal a more complex pattern (left inset) that does not seem to be compatible with the simple star phylogeny observed in the example from \FIG{HIV_example2}A. However, when we focus on the two most abundant haplotypes 0 and 1, we see that these two haplotypes differ at more than one site from each other. The majority of the remaining haplotypes differs by only one mutation from either haplotype 0 or 1. This pattern can be interpreted as the result of two independent partial sweeps where the two founding haplotypes (0 and 1) differ at two sites.

We can disentangle these two hard sweeps by assigning the rare haplotypes to either of the two founding haplotypes if they from the latter by only one mutation. When reordering the distance matrix according to this assignment, we recover the two individual sweep pattern as blocks on the diagonal. This is shown in the right inset of~\FIG{HIV_example2}B. Both reordered components now exhibit the characteristic power-law decay in their individual rank-frequency spectra with similar selection coefficients, again in the range of a few percent.

A scenario as the one shown in~\FIG{HIV_example2}B, with several haplotypes carrying the same adaptive mutation rising in frequency simultaneously, is commonly referred to as a ``soft sweep''~\citep{Hermisson:2005p1649}. Soft sweeps can occur, for example, if an adaptive mutation arises repeatedly on different haplotypes, which is expected to be common in viruses with large population sizes and high mutation rates such as HIV, where every single point mutation can arise many times each generation. Alternatively, soft sweeps can result when a sudden change in environment renders a previously neutral or deleterious allele selectively beneficial and adaptation involves several different adaptive haplotypes from the standing genetic variation.

An unambiguous decomposition of the haplotype distance-matrix into individual sweep components of a soft sweep requires that the founding haplotypes differ by several mutations. This is likely to be the case for a diverse ancestral population. Note that the above approach also suggests how $\hat{s}$ can be applied to incomplete sweeps or sweeps restricted to a subpopulation. In those cases, the reordered distance matrix should display the characteristic sweep pattern only for a sub-block on the diagonal that is embedded into the diverse remainder of the population.

Intriguingly, we see very little evidence of degradation of the haplotype spectrum by genetic drift. This might have several causes: (i) a very large population size, (ii) a very recent sweep, or (iii) a scenario where sweeps are so frequent that they start overlapping such that exponential amplification of the most fit variants never ceases. The latter is expected in large continuously adapting populations \citep{Neher:2011p42539}. To investigate this matter further, one would need denser time-course data and information on genetic diversity across the genome.

\section*{Discussion}

We have investigated the pattern of haplotype variation that arises from mutations occurring during the early phase of a selective sweep on the sweeping haplotype. We found that a selective sweep leaves a characteristic signature in the frequency spectrum of such novel haplotypes: when ordering all adaptive haplotype variants by their abundance, the frequency of the $i$th haplotype variant is approximately $\mut/(is)$ times the frequency of the first, where $\mut$ is the mutation rate of the locus and $s$ is the selection coefficient of the sweep. The power-law decay of the rank-frequency spectrum is a consequence of exponential amplification (via selection) competing with diversification (through mutation and recombination), similar to the mechanism of preferential attachment originally proposed by~\citet{Yule:1925p38579}. We applied this finding to construct an estimator for the selection coefficient at a candidate locus where a selective sweep is suspected to have occurred recently. Such loci could be either identified by standard approaches to localize sweeps~\citep{Fay:2000p35077,Sabeti:2002p18853,Kim:2002p37296,Nielsen:2005p42523, Voight:2006p18353,Sabeti:2007p12131}, or a priori information could suggest that a sweep has occurred, such as in our HIV examples, where the failure of anti-retroviral therapy implied an adaptive change at the RT locus.

The classic signature used to infer the strength of a recent sweep has been the size of the dip in diversity around the adaptive site~\citep{Smith:1974p34217}. Underlying this approach is the rationale that recombination breaks up linkage with increasing genetic distance from the adaptive site: The further away from the sweeping locus, the higher the probability that pre-sweep variation has become unlinked from the adaptive allele and thus remains polymorphic after the sweep. Hence, the width of the dip in genetic diversity scales with the ratio of the selection coefficient and the recombination rate.

While successfully applied in many studies~\citep{Kim:2002p37296,Andolfatto:2007p18246,Macpherson:2007p18421,Sabeti:2002p18853,Sabeti:2007p12131,Sattath:2011p42514}, this approach suffers from a number of shortcomings that can limit its applicability. First, it relies on recombination and thus cannot be applied in organisms which frequently self, such as many plants or yeast, as well as organisms that recombine via horiontal gene transfer, such as bacteria. Recombination rates can also fluctuate strongly along the genome and in time~\citep{Winckler:2005p11216,Coop:2007p42685}, rendering their precise estimation difficult, especially in the regions of reduced genetic diversity around a selective sweep. Furthermore, the approach relies on the presence of pre-existing variation at the sweep locus and assumes that we know how much variation was present originally. Ancestral diversity is literally absent in evolution experiments where populations start from a single clone, or in populations where adaptation occurs so frequently that genetic diversity is not fully restored between recurrent selective sweeps~\citep{Gillespie:1994p42649}. Finally, for strong sweeps, the dips in diversity can potentially span very large regions extending up to entire chromosomes \citep{Andersen:2012p45984}. In such cases, dip-sizes can no longer be assessed accurately.

Our estimator is less dependent on ancestral diversity or accurate recombination rates estimates, since we investigate the new variation that emerges during the sweep and compare selection strength to rate of haplotype diversification (mutation and recombination) rather than to the recombination rate alone. It is essential for our estimator that one can accurately determine haplotype population frequencies on the order of $\mut/(s\is)$, requiring deep population samples. For example, if we assume a sweep locus with $\mut/s = 0.1$ and base our analysis on the five most frequent adaptive haplotype variants ($\is=5$), it would be necessary to measure haplotype frequencies of~2\% accurately. This calls for a sample size of roughly~$10^3$. Population genetic data of such depth is already available for HIV, where a large number of sequences can be obtained from plasma samples of infected patients \citep{Tsibris:2009p26554,Fischer:2010p40314,Hedskog:2010p36144}. Several efforts are currently being made to achieve a comparable in-depth characterization of the population diversity in several eukaryotes including humans~\citep{1000GenomesProjectConsortium:2010p37747}, flies~\citep{DPGP}, and plants~\citep{Cao:2011p43483}.

In practice, one typically has a data set (as in the HIV example a sample of size $\approx 10^4$ sequences of length 120-200bp) and wants to estimate $s$. The size of the locus corresponds to a certain mutation rate, which will limit the range of selection coefficients that give rise to an observable sweep spectrum: If $u/s$ is too small (say smaller than 10 times the inverse sample size), very little variation will be observed. On the other hand, if $u/s$ is close to one, the founding haplotype diversifies as rapidly as it is amplified and will no longer be the dominant haplotype in the sample. Without a dominant haplotype, the method cannot be applied. Furthermore, our calculations always assume that $u\ll s$.

Too large values of $u/s$ can be circumvented by restricting the analysis to only a fraction of the locus, which effectively reduces $u$. Hence, if a sample contains a large number of rare alleles and haplotypes but no dominant haplotype, one can reduce the window size to see whether a dominant haplotype with a trailing sweep spectrum emerges. Similarly, if one analyzes long genomes using a windowing approach and observes a region almost devoid of diversity (corresponding to $u/s \lll 1$), one can increase $u$ by increasing the window size until several low frequency haplotypes are observed. One can thus tune the sensitivity of the estimator to different ranges of $s$ by adjusting $u$ through the length of the locus. A selection coefficient of $s=10^{-3}$ and a mutation rate of $10^{-8}$ per site, for example, would require a locus of length 10 kb to achieve $u/s=0.1$.

We used the popular program \texttt{sweepfinder} \citep{Nielsen:2005p42523} to compare the performance of our estimator to a traditional approach where the selection coefficient of a sweep is inferred from its signature in the surrounding ancestral diversity. As expected, our estimator consistently outperforms the traditional approach if ancestral diversity is low (Supplementary Fig. 2). Interestingly, even when ancestral diversity is rather high (e.g. $\Theta=0.01$, comparable to the level of neutral diversity observed in \emph{D. melanogaster}), the estimates from our method still have substantially lower variance than those obtained from  \texttt{sweepfinder}. Note, however, that the comparison between the two methods is based on rather different data. While the  \texttt{sweepfinder} requires long sequences ($r/s \approx 1$) at moderate coverage, our methods works with much shorter regions ($(\mut+r)/s <0.1$) at very deep coverage.

In order for our estimator to be applicable, haplotype variants that were produced after a sweep and rose in frequency by random genetic drift need to be still at low-enough frequency such that they have not yet degraded the sweep signature. The extent of this degradation can be estimated from a simple argument: the lowest population frequency entering our estimator is of the order $\mut/(s\is)$, while drift-dominated haplotypes will typically be at frequencies $t/N$, where $t$ is the time that has elapsed since the sweep. We thus require $\mut/(s\is)\gg t/N$, which translates into the condition that population sizes have to be sufficiently large and that sweeps are not too old. As shown in \FIG{sweep_spectra_relaxation}, drift results in a growing bulge at low frequencies where the rank spectrum is exponential rather than a power-law. Since the strength of genetic drift is very poorly known, we propose to inspect the ranked haplotype spectrum for deviations from the power-law and choose $\is$ small enough such that drift dominated haplotypes are excluded from the analysis. In addition, one should check whether the sample is compatible with a near star phylogeny. Failure to exhibit these features should indicate either the absence of a selective sweep, or a sweep that has already been degraded.

While we have focussed on mutation as a source of novel haplotypes, recombination can produce new haplotypes as well. Provided each recombination event results in a unique haplotype, all of the above formulae hold with $\mut+r$ instead of $\mut$ alone. If, however, recombination rates are not known, recombinant haplotypes can be filtered out by restricting the analysis to only those haplotypes that differ by a single mutation from the founding haplotype. This removes the majority of recombinant haplotypes since recombination with the diverse ancestral population typically incorporates several polymorphisms at once. 

Recent studies suggest that in many selective sweeps the adaptive allele has not actually become fixed in the population (incomplete sweep), or that several different haplotypes, all carrying the adaptive allele, have swept simultaneously through the population (soft sweep)~\citep{Hermisson:2005p1649,Pritchard:2010p35368,Karasov:2010p35377,Burke:2010p42546}. We have demonstrated how incomplete sweeps and soft sweeps can be analyzed by our method (\FIG{HIV_example2}B), making use of the fact that the novel haplotype variants our estimator is based on are related to the founding haplotype by a simple star phylogeny, which allows to easily differentiate them from other haplotypes that are not descendants of the founding haplotype.

Our results on the haplotype pattern of a selective sweep were derived under the assumption of a panmictic population of constant size, raising the question how they might be affected by past demographic events such as population expansions or bottlenecks. Most current approaches to investigate selective sweeps rely on the specific patterns a sweep leaves in ancestral neutral diversity. These approaches can thus be very sensitive to past demographic events that shape the patterns of ancestral diversity over a time scale of neutral coalescence, which can include quite ancient demographic events. The approach presented here, however, is fundamentally different. In contrast to ancestral neutral variation, which is typically old, we focus on the very recent variation that arises during the early phase of a selective sweep. Demographic events that occurred prior to the onset of the sweep are thus irrelevant. Only very rapid changes in population size that happen while the adaptive allele is sweeping can cause significant deviations in the haplotype frequency spectra.

The key scenario to be discussed in this context is that of a recent population expansion, since our estimator measures specifically the rate at which new haplotype variants were amplified. Indeed, the haplotype frequency spectrum in an expanding population will resemble that of a selective sweep if the expansion lasted long enough for all coalescence to happen during the expansion and if the expansion rate $\epsilon$ is large enough such that the spectrum has not yet been eroded by drift, i.e., $N\epsilon\gg 1$. In this case, our estimator turns into an estimator of the expansion rate that might be applicable to scenarios such as the expansion of an HIV population in a newly infected individual or the spread of novel strains of influenza. If, on top of an expansion, a beneficial mutation is spreading through the population, the haplotype carrying the beneficial mutation (and its descendants) is expanding faster than the ancestral haplotypes. The estimate of the selection coefficient will then in fact be an estimate of $s+\epsilon$.

The spread of a beneficial mutation in the population generally reduces genetic diversity in the vicinity of the adaptive site. That a selective sweep can also amplify new diversity at very low population frequencies is thereby often overlooked. We have shown that the spectrum of this new variation records the exponential amplification of the novel beneficial allele  in a clock-like fashion and can thus be used to estimate its selection coefficient. With the recent advances in sequencing technologies, the required information about low-frequency genetic variation is no longer elusive, making our estimator applicable for a wide range of analyses.

\section*{Acknowledgments}
This research was supported by the European Research Council under Grant No.~260686 (RAN). PWM was an HFSP postdoctoral fellow in the Petrov lab at Stanford University. Part of this work was conducted during the program ``Microbial and Viral Evolution'' at KITP supported by the National Science Foundation under Grant No. PHY05-51164. We thank Dmitri Petrov, Nick Barton, Boris Shraiman, Ben Callahan, Taylor Kessinger, and members of the Petrov lab for helpful feedback.

\newpage
\appendix

\section{Neutral haplotype spectrum}
\label{sec:app_neutral_spectrum}
The haplotype spectrum expected in a haploid neutral Fisher-Wright model without recombination can be calculated from the Ewens sampling formula \citep{Ewens:1972p35481}. Ewens showed that the probability of a sample of size $n$ is
\begin{equation}
P_n(a_1, a_2,\ldots, a_n) = \frac{n!}{\Theta_{(n)}}\prod_{m=1}^n
\left(\frac{\Theta}{m}\right)^{a_m}\frac{1}{a_m!} \ ,
\end{equation}
where $a_j$ is the number of allele classes that are sampled $j$ times and
$\Theta_{(k)}= \Theta(\Theta+1)\cdots(\Theta+k-1)$ with $\Theta=2N\mut$. The expectation of
$a_k$ is therefore given by
\begin{equation}
\begin{split}
\langle a_k \rangle &= \sum_{\bf{a}}
\frac{\Theta}{k}
\left(\frac{\Theta}{k}\right)^{a_k-1}\frac{1}{(a_k-1)!}
\frac{n!}{\Theta_{(n)}}\prod_{m\neq k}
\left(\frac{\Theta}{m}\right)^{a_m}\frac{1}{a_m!}\\
&=\sum_{\bf{a}}
\frac{\Theta \Theta_{(n-k)}n!}{k \Theta_{(n)}(n-k)!}
P_{n-k}(a_1, a_2,\ldots, a_n) = \frac{\Theta \Theta_{(n-k)}n!}{k
\Theta_{(n)}(n-k)!} \\
& =\frac{\Theta}{k} \prod_{m=n-k+1}^n \frac{m}{\Theta + m -1}
=\frac{\Theta}{k} \prod_{m=n-k+1}^n \left(1+\frac{\Theta-1}{m}\right)^{-1}
\approx \frac{\Theta}{k} e^{-(\Theta -1) k/n} \approx \frac{\Theta}{k}
\end{split}
\end{equation}
where the last two approximate inequalities are accurate if $k\ll n$ and $k\Theta \ll n$, respectively.
Hence the expected number $\is$ of allele classes with more than $\ns$ members
is roughly $\Theta \sum_{k=\ns}^{n\Theta^{-1}} k^{-1} \approx
\Theta (\log n- \log \Theta\ns)$, where cutting off the sum at $k=n\Theta^{-1}$ approximately accounts for the exponential. With this approximation, the $\is$th abundant allele class is expected to contain
\begin{equation}
\ns \approx \frac{n}{\Theta} \exp(-\is/\Theta)
\end{equation}
copies of the allele. A more accurate expression of the spectrum is obtained by determining the $\ns$ such that $\is = \sum_{k>\ns} \langle a_k \rangle$, using the exact expression given above. This numerical solution for the haplotype spectrum is plotted in Figure 2B of the main text.

\section{The distribution of haplotype frequencies}
\label{sec:joint_distribution}
In the main text, we calculated the distribution of the establishment time of the $i$th haplotype and the frequency of the corresponding haplotype. Here, we show how the joint distribution of all seeding times and the resulting frequency spectrum can be calculated assuming that the novel haplotypes are rare and evolve independently, which is justified if they constitute a small share of the total population, i.e., if $\mut/s\ll1$. In this case, the probability that $k$ haplotypes $i=1,\ldots, k$ are present in frequencies $x_i$ is given by
\begin{equation}
\label{eq:joint_dis}
P(x_1,\ldots, x_k|t) = \int_0^{t} \prod_i dt_i \prod_i P(x_i | t_i,t) P(t_1\ldots, t_k|t)\ ,
\end{equation}
where $P(x_i | t_i,t)$ is the probability that a haplotype has frequency $x_i$ at time $t$ given it became established at time $t_i$. The distribution of establishment times $P(t_1\ldots, t_k|t)$ is given by
\begin{equation}
\label{eq:est_time_dis}
P(t_1\ldots, t_k |t) = \frac{1}{k!}e^{-\int_0^t dt' \alpha(t')} \prod_i \alpha(t_i)\ ,
\end{equation}
where $\alpha(t')=2s\mut N x(t')$ is the rate of establishing novel adaptive haplotypes (main text Equation (1) and below).
Note that the $t_i$ defined in \EQ{est_time_dis} are not ordered. They are distributed according to a Poisson point process with density $\alpha(t')$. Assuming that established novel haplotypes increase in frequency logistically according to Equation (5) of the main text, we have
\begin{equation}
\label{eq:phap}
P(x_i | t_i,t) = \delta\left(x_i - \frac{e^{(s-\mut)(t-t_i)}}{2Ns + e^{st}}\right)\ ,
\end{equation}
where $\delta(x)$ is the Dirac $\delta$-function  (the stochastic analog is calculated below, see also \citep{Desai:2007p954}). Substituting $P(x_i | t_i,t)$ into \EQ{joint_dis} and integrating over $t_i$, we obtain
\begin{equation}
P(x_1,\ldots, x_k|t) = \frac{1}{k!}e^{-\int_0^t dt' \alpha(t')} \prod_i \frac{1}{(s-\mut)x_i} \alpha(t_i)
\end{equation}
with $t_i = t-(s-\mut)^{-1} \log (x_i(2Ns-e^{st}))$.
Haplotypes that are common after the sweep are most likely seeded early during the sweep. Furthermore, we showed in Equation (5) of the main text that their relative frequencies stay approximately constant during the amplification phase. Hence we can determine the joint distribution of frequencies at early times $t\ll s^{-1}\log 2Ns$ while $\alpha(t) \approx 2 \mut s e^{st}$ is still exponential. After substituting the $t_i$ and simplifying, we find
\begin{equation}
\begin{split}
P(x_1,\ldots, x_k|t) &\approx \frac{e^{-\frac{\mut}{s}e^{st}}}{k!}\prod_i \frac{2 \mut s e^{s t_i}}{(s-\mut)x_i}
\sim \prod_i x_i^{-2-\frac{\mut}{s}} \ ,
\end{split}
\end{equation}
where we dropped factors independent of $x_i$ which ensure normalization. A very similar result was found in \citep{Desai:2007p954}. At large $t$, the form of the prefactor $e^{-\frac{\mut}{s}e^{st}}$ changes due to the saturation of the allele frequency at $1$, but the distribution of the frequencies of the haplotypes that were seeded early during the sweep remains of this form until the spectrum is eroded by genetic drift.

The haplotype spectrum therefore decays with a power $2+\mut/s$, which is consistent with the power $1-\mut/s$ obtained for the cumulative or rank spectrum (integrating $x_i^{-2-\mut/s}$ yields $x_i^{-1-\mut/s}$). More importantly, this result tells us that the distribution of haplotype frequencies conditional on the number of haplotypes observed is approximately independent of $\mut/s$ if $\mut\ll s$. Hence, given that a sweep occurred, all information about the strength of the sweep is contained in the number of haplotypes and the precise values of their frequencies do not contain any additional information if $\mut\ll s$. However, whenever there are deviations from the assumptions made here, the haplotype frequencies will contain additional information.

\section{Stochastic derivation of the haplotype spectrum}
The dynamics of rare haplotypes are strongly influenced by random genetic drift and we have to ascertain the deterministic arguments made in the main text by a more careful stochastic calculation. While hard in general, an approximate analytic calculation of the frequency spectrum of rare haplotypes is feasible in our case for the following reasons: (i) The dynamics of a beneficial allele are essentially deterministic since it is much more frequent than haplotypes that arise through secondary mutations. (ii) The dynamics of rare haplotypes can be described by a linear branching process since they are always a small fraction of the total population.

As already done in \EQ{joint_dis}, we decompose the distribution of haplotype frequencies into the distribution $P(t_1,\ldots,t_k|t)$  of times when the novel haplotypes arise and probability $P(n, t|t_0)$ that a haplotype is present in $n$ copies at time $t$, given it arose at time $t_0$. We will derive $P(n, t|t_0)$ first and consider the spectrum due to the superposition of several independent seeding events below.

\subsection{Distribution of rare variants arising in a logistic sweep}
\label{sec:app_distribution}
To model the stochastic dynamics of rare haplotypes, we use a continuous time branching process in which individuals produce identical copies of themselves with rate $1+\gr(t)$ and die with rate $1$, i.e., the unit of time is chosen to be the generation time. The average number of offspring of a given individual in this model is $1+\gr(t)$. Hence, $\gr(t)$ is the growth rate of the haplotype carrying the beneficial allele. In the case of a sweep, we have
\begin{equation}
\gr(t) = s(1-x(t)) -\mut \ ,
\end{equation}
where the first term accounts for selection ($x(t)$ is the frequency of the beneficial allele) and the second term accounts for mutations that change the state of the haplotype. The dynamics of $P(n,t| t_0)$ are described by the forward Master equation
\begin{equation}
\partial_t P(n,t|t_0) = (1+\gr(t))(n-1)P(n-1,t|t_0)+(n+1)P(n+1,t|t_0) -(2+\gr(t)) n P(n,t|t_0) \ ,
\end{equation}
which accounts for replication (first term) and death (second term).
To solve for $P(n,t|t_0)$, it is useful to consider the generating function $G(\lambda,t|t_0) = \sum \lambda^n P(n,t|t_0)$, which obeys the equation
\begin{equation}
\partial_t G(\lambda,t|t_0) = \left[-(2+\gr(t)) \lambda  +
(1+\gr(t))\lambda^2 +1\right]\partial_\lambda
G(\lambda,t|t_0)
\end{equation}
with initial condition $G(\lambda, t_0|t_0)=\lambda$. This equation can be solved via the method of characteristics, with the result
\begin{equation}
\label{eq:generating_function}
G(\lambda, t|t_0) \approx  1- \frac{1-\lambda}{e^{-\int_{t_0}^t dt'
\gr(t')}+(1-\lambda) \int_{t_0}^{t} dt' e^{-\int_{t_0}^{t'} dt'' \gr(t'')}} \ ,
\end{equation}
where we have used $1+\gr(t)\approx 1$ along the way. The latter is a good approximation if selection is weak in one generation and amounts to neglecting terms of order $s^2$.  We will now substitute the explicit expression for $\gr(t)$, where it will be convenient to parametrize the frequency of the beneficial allele as $x(t) = (1+e^{s(\tau-t)})^{-1}$ with $\tau = s^{-1}\log 2Ns$.
Using this form of $\gr(t)$, we find for the generating function
\begin{equation}
G(\lambda, t|t_0)=1- \frac{\su(1-\lambda)(1+e^{-s(t_0-\tau)})e^{-\mut(t-t_0)}}{\su +
\su e^{-s(t-\tau)}+(1-\lambda) \left[e^{-s(t_0-\tau)-\mut(t-t_0)}-e^{-s (t-\tau)}+\su\mut^{-1} (1-e^{-\mut(t-t_0)})\right]}
\end{equation}
where $\su=s-\mut$. Any haplotype that is abundant enough to be sampled with high probability most likely originated in the early phase of the sweep ($t_0\ll \tau$), which allows for the approximation $1+e^{-s(t_0-\tau)}\approx e^{-s(t_0-\tau)}(1+\mathcal{O}(n/(sN)))$ where $n$ is the sample size. Furthermore, we will typically observe the spectrum at times $t\gg \tau$ when the sweep is almost complete. Hence we can approximate $1+e^{-s(t-\tau)}\approx 1 + x_{WT}\approx 1$ where $x_{WT}$ is the frequency of the deleterious wild type allele at the time of sampling.
Using these simplifications, we obtain
\begin{equation}
G(\lambda, t|t_0)\approx 1- \frac{(1-\lambda)e^{-s(t_0-\tau)-\mut(t-t_0)}}{1
+(1-\lambda) (\su^{-1}e^{-s(t_0-\tau)-\mut(t-t_0)}+ \mut^{-1}(1-e^{-\mut(t-t_0)}))}
\end{equation}
This expression is straightforwardly expanded into a geometric series in $\lambda$ whose
coefficients are $P(n,t|t_0)$. For large $n$, one
finds
\begin{equation}
P(n,t|t_0) \approx
\frac{e^{-s(t_0-\tau)-u(t-t_0)}}{\nref^2}e^{-n/\nref}\quad
\mathrm{where}\quad \nref =\frac{e^{-s(t_0-\tau)-u(t-t_0)}}{s-u}+ \frac{1-e^{-u(t-t_0)}}{u} \ ,
\end{equation}
with relative corrections being on the order of $\nref^{-1}$.
The quantity $\nref$ is the mean copy number of the haplotype conditional on non-extinction, and the two terms contributing to $\nref$ have a straightforward interpretation: The first term is the contribution of selection, which amplifies the haplotype before the fixation of the beneficial allele. The second term is the contribution of random genetic drift, which evaluates simply to $t-t_0$ in the limit of $\mut(t-t_0)\ll 1$. The latter is the analog of the well known fact that a non-extinct neutral allele in a neutral Moran process is on average present in $n$ copies after $n$ generations. The expression for $\nref$ exhibits a crossover from an early regime where selection dominates $\nref$ to a random drift dominated regime at large $t$. In the limit $\mut(t-t_0)\ll 1$, we have
\begin{equation}
\nref(t)\approx \begin{cases}
   (s-\mut)^{-1}e^{s(\tau-t_0)} & s(t-t_0)\ll e^{s(\tau-t_0)}\\
   (t-t_0) & s(t-t_0)\gg e^{s(\tau-t_0)}
   \end{cases}
\end{equation}
This crossover will inform us below about how long the contribution of random drift can be neglected when applying our estimator of the strength of the selective sweep.

\subsection{The haplotype frequency spectrum}
\label{sec:app_superposition}
Having calculated the copy number distribution of a haplotype that originated at time $t_0$, we now have to determine the distribution of seeding times and calculate the resulting spectrum of haplotype frequencies.
New haplotypes that contain the beneficial allele are produced at rate
\begin{equation}
\gamma(t) = N\mut x(t) = \frac{N\mut}{1+e^{-s(t-\tau)}} \ ,
\end{equation}
where, as before, $x(t)$ is the frequency of the sweeping allele. Note that this differs from the rate of establishment  of novel variants by a factor $s$, which will reemerge from the stochastic calculation. The deterministic approximation for $\gamma(t)$ is valid if it is unlikely that new variants are seeded before establishment of the founding variant, which requires $s\gg \mut$ (see \citet{Desai:2007p954}). Since novel haplotypes are seeded and evolve independently to a good approximation, the number of haplotypes present in $n$ copies at time $t$ is Poisson distributed with mean
\begin{equation}
Q(n, t) = \int_0^t dt_0\; \gamma(t_0) P(n,t|t_0) \ ,
\end{equation}
Due to the exponential nature of $P(n,t|t_0)$, it is convenient to sum $Q(n,t)$ over $n>\ns$ and calculate the expected number of haplotypes with copy numbers greater than $\ns$. The sum is well approximated by the integral $W(\ns, t) = \int_{\ns} dn Q(n,t)$, and we have
\begin{equation}
\begin{split}
\label{eq:app_W}
W(\ns, t) &= N\mut \int_{\ns}^\infty dn \int_0^t dt_0 x(t_0) P(n,t|t_0) \\
&=N\mut \int_0^t dt_0 \frac{e^{-s(t_0-\tau)-u(t-t_0)}}{\nref(1+e^{-s(t_0-\tau)})} e^{-\ns/\nref}
\approx N\mut \int_0^t dt_0 \frac{e^{-u(t-t_0)}}{\nref} e^{-\ns/\nref} \ ,
\end{split}
\end{equation}
where the last approximation assumes that novel haplotypes are seeded while the beneficial allele is still expanding exponentially, i.e., $e^{-s(t_0-\tau)}\gg 1$.

It does not seem possible to evaluate the integral over $t_0$ in \EQ{app_W} analytically. However, the integral is dominated by contributions from a well defined time interval and can  be evaluated perturbatively. For  a very large population where drift is negligible, and for $s\gg \mut$, this integral simplifies to
\begin{equation}
W(\ns,t)\approx N\mut \int_0^t dt_0 e^{s(t_0-\tau)-s\ns e^{s(t_0-\tau)}} \ ,
\end{equation}
which is very sharply cutoff for $s(t_0-\tau) \gg -\log s\ns$. Hence we can send the upper integration boundary to infinity without loss of accuracy and evaluate this integral exactly. One finds $W(\ns,t)\approx N\mut/(s\ns)$. The contribution to this integral come from a narrow peak of width $s^{-1}$ and height $N\mut e^{-1}/\ns$. Genetic drift and mutation predominantly change only this height and width, leaving the shape of the integral approximately invariant. Hence we can evaluate this integral by calculating where the integral peaks and how wide this peak is (Laplace's method).

Including the correction due to drift and finite $s/\mut$ term corrections, the integrand peaks when
$\ns \approx \nref$, which translates into $\su(\tau-t^*) \approx \log (s(\ns-t+t^*))$ as opposed to $s(\tau-t^*) \approx \log s\ns$ without the corrections. The second derivative of the logarithm of the integrand is approximately given approximately by
\begin{equation}
\frac{1}{\nref^2}\left(\frac{d\nref}{d t_0}\right)^2 \approx s^2\frac{(\nref - \Delta t)^2}{\nref^2} \ ,
\end{equation}
where $\Delta t =t-t^*$ is the age of the haplotypes. Hence the peak dominating the integral becomes wider by a factor $\frac{\nref}{\nref-\Delta t}$.
With these corrections to the ``height'' and the ``width'' of the integrand, we obtain the approximate expression
\begin{equation}
\label{eq:approx_number_ofhap}
W(\ns,t)\approx \frac{N\mut e^{-\mut t}}{s(\ns-\Delta t)}\left(\frac{N}{\ns}\right)^{u/s}
\end{equation}
for the integral. This expression is accurate as long as $s\ns\gg s(t-\tau)-\log(s\ns)$ and $\mut t\ll 1$. The age of the haplotype evaluates approximately to $\Delta t=t-\tau+s^{-1}\log(s\ns)$, which is of order $\tau$.
The additional factor $(N/\ns)^{\mut/s}$ accounts for the additional time the older haplotypes have been degraded by mutations, while the $\ns - \Delta t$ accounts for the contribution of drift to the frequency of the haplotypes.

After the sweep, the frequency of the most common haplotype is $x_0=e^{-\mut t}$ and the expected number of haplotypes above frequency $\xs$ is given by
\begin{equation}
\label{eq:approx_number_ofhap_above_freq}
W(\xs, t) \approx
\frac{\mut}{s}\left[\frac{x_0}{(\xs-\Delta t/N)}\right]^{1+\mut/s} \ .
\end{equation}
This result tells us that the mean number of haplotypes with frequencies greater than $\xs$ is approximately linear in $\mut/s$ and decreases with $\xs$ approximately as $\xs^{-1}$. Furthermore, the expected number of haplotypes above $\xs$ is increasing with time, since rare haplotypes increase in frequency due to random drift.

Given that we observe $\is$ haplotypes at frequencies higher then $\xs$, we can use \EQ{approx_number_ofhap_above_freq} to estimate $s/\mut$:
\begin{equation}
\label{eq:app_estimate}
\frac{s}{\mut} \approx
\frac{1}{\is} \left[\frac{x_0}{(\xs-\Delta t/N)}\right]^{1+\is\xs/x_0} \ ,
\end{equation}
This equation differs from Equation (9) of the main text by a reduction of $\xs$ due to random drift, which has been ignored in the simple deterministic derivation given in the main text. This reduction allows us to correct for the effects of genetic drift as long as drift is not to strong. Obviously, the correction fails as $\Delta t$ approaches $N\xs$.

\EQ{app_estimate} informs us about the regimes where the proposed methods to
estimate the selection coefficient is likely to work. Random genetic drift will degrade the signature of the sweep for haplotype frequencies smaller than $\Delta t/N$. Since the time needed for completion of the sweep is on the order of $(\log Ns)/s$, we require $N s \xs > \log N s$. Since $\xs \approx \mut/(s\is)$ is itself small, we need  $N s \gg \is\log N s$ for the method to work. The breakdown of the method is clearly seen in Figure 3 of the main text once $Ns$ falls below 100.

\section{Pruning recombinant haplotypes}
\label{sec:app_pruning}
Haplotypes that arise by mutation from the founding haplotype differ at exactly one position from the founding haplotype, while haplotypes that result from a recombination event with a member of the diverse ancestral population typically differ at several positions. Furthermore, haplotypes that are mutants of mutants will differ at two positions from the founding haplotype. In the main text, we argued that one can restrict the haplotype spectrum to those haplotypes that differ only at a single site from the founding haplotype, thereby removing most recombinant haplotypes and mutants of mutants. Here, we show that frequency spectrum of such a restricted set of haplotypes differs slightly from that of all haplotypes.

As before, the frequency of the beneficial allele will typically follow Equation (1) of the main text. The frequency of the founding haplotype, however, will remain below this frequency due to loss through recombination and mutation.
\begin{equation}
x_0(t) = \frac{e^{\sur t}}{2Ns + e^{st}} \ ,
\end{equation}
where we have abbreviated the initial growth rate of a haplotype by $\sur =s-\mut-r$.
Mutations on this haplotype establish with rate $\alpha(t)=2\mut \sur Nx_0(t)$. Haplotypes that establish at time $t_i$  then typically follow a frequency trajectory
\begin{equation}
x_i(t) = \frac{e^{\sur (t-t_i)}}{2N\sur  + e^{st}} \ .
\end{equation}
The most likely seeding time of the ith haplotype is given by
\begin{equation}
t_i = \frac{1}{\sur}\log \frac{s i}{\mut} \ .
\end{equation}
Hence we obtain for the ratios of haplotype frequencies at times when the beneficial allele is near fixation
\begin{equation}
\frac{x_i(t)}{x_0(t)} = e^{-\sur t_i} = \frac{\mut}{s i} \ .
\end{equation}
This differs from Equation (7) of the main text in that the ratio is proportional to $i^{-1}$, rather than $i^{-1+\mut/s}$. This difference is due to the fact that here haplotypes grow with the same rate as the rate at which they are seeded. In the previous case where all haplotypes are considered,  haplotypes grow with rate $s-\mut-r$, while the seeding rate is proportional to the frequency of the beneficial allele which grows with rate $s$.

\begin{figure}[htp]
\begin{center}
\includegraphics[width=0.49\columnwidth,type=pdf,ext=.pdf,read=.pdf]{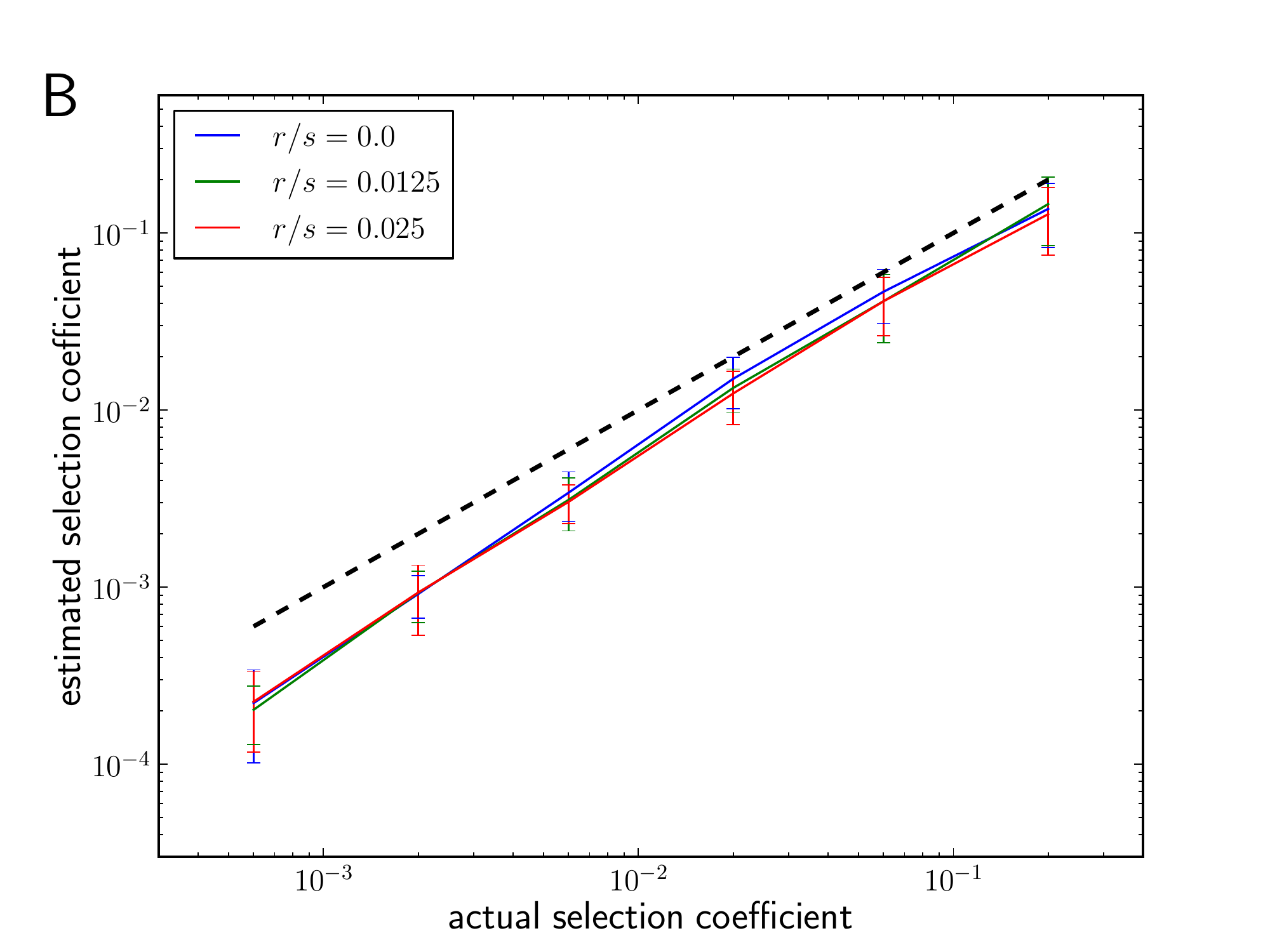}
\includegraphics[width=0.49\columnwidth,type=pdf,ext=.pdf,read=.pdf]{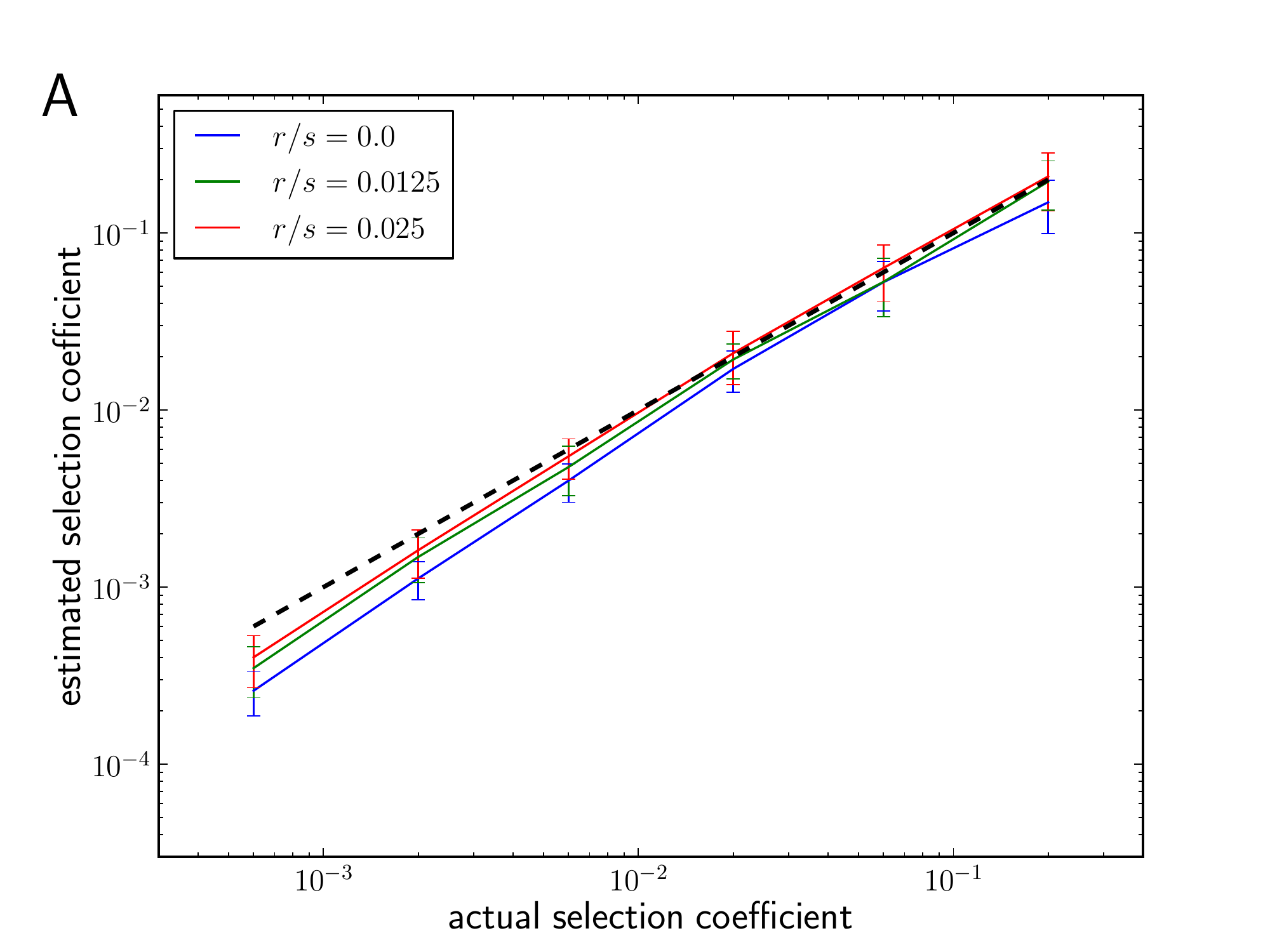}  
\caption[labelInTOC]{Estimating selection strength in presence of recombination. This figure is the analog of Fig 4 of the main text with a ten-fold reduced number of segregating sites in the initial samples, i.e., lower ancestral diversity. The estimates are very similar, showing that the ancestral diversity has no impact on the accuracy of the estimation as long as recombination events almost always give rise to unique haplotypes that differ from previous haplotypes at more than 1 site. }
 \label{fig:supp_recombination}
\end{center}
\end{figure}

\newpage

\begin{figure*}[htp]
\begin{center}
\includegraphics[width=0.49\columnwidth,type=pdf,ext=.pdf,read=.pdf]{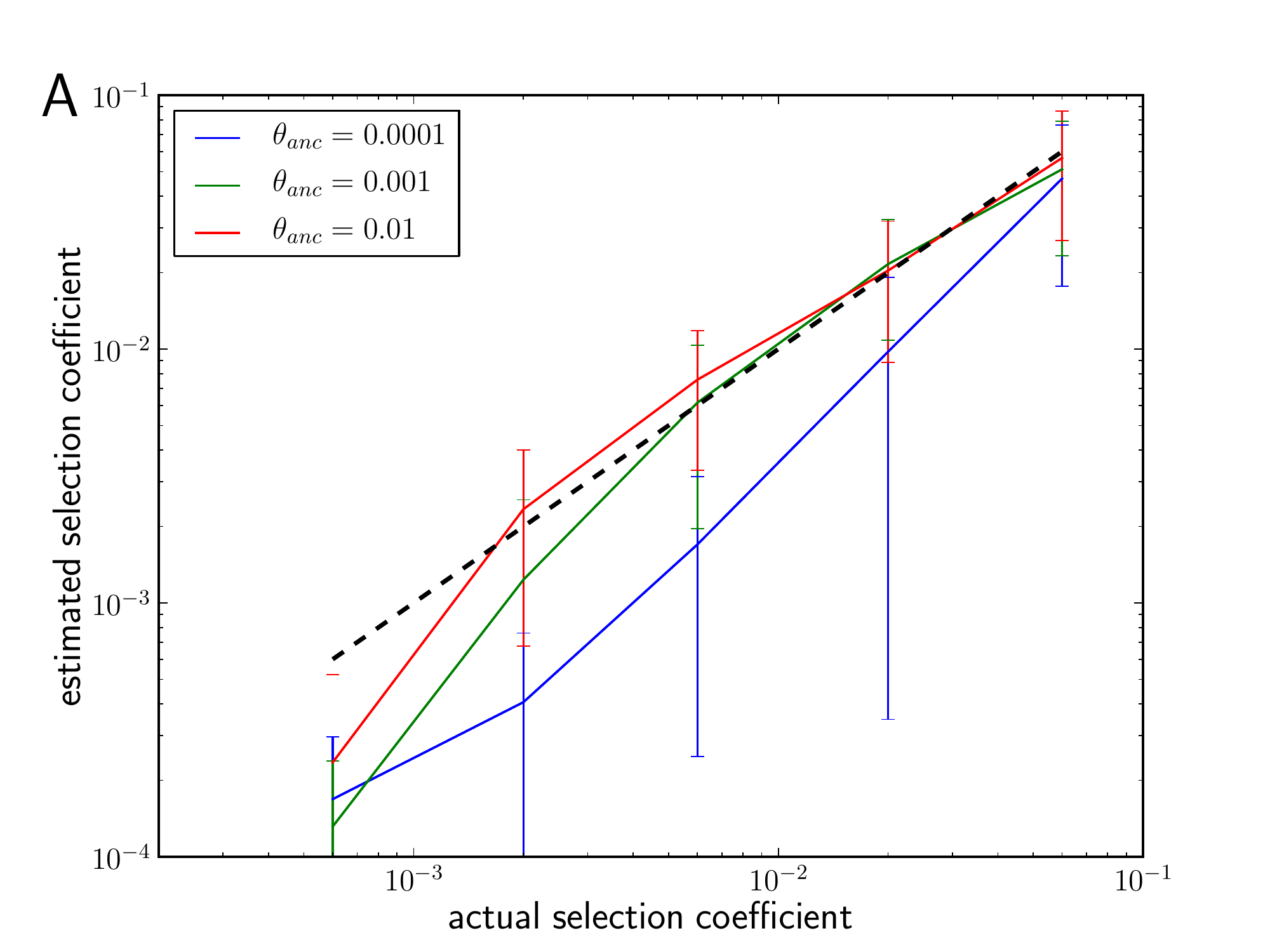}
\includegraphics[width=0.49\columnwidth,type=pdf,ext=.pdf,read=.pdf]{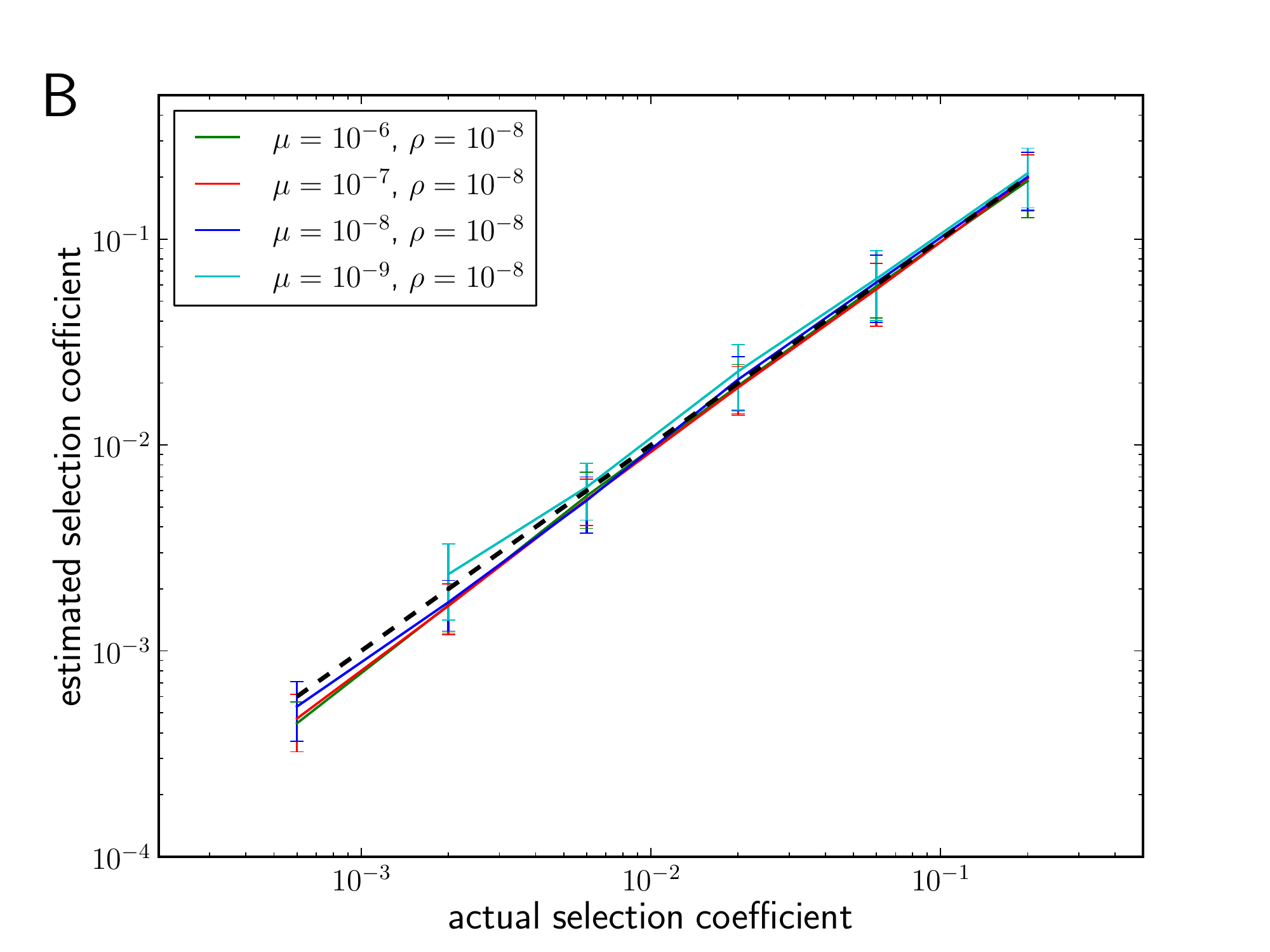}  
\caption[labelInTOC]{Comparison of the accuracy of our estimator to the accuracy of estimates obtained from the program \texttt{sweepfinder} \citep{Nielsen:2005p42523}. Haplotype samples around a selective sweep occurring in the middle of the locus were simulated with the program \texttt{msms} \citep{Ewing:2010p43021}. The population size used was $N=10^5$ and samples were taken when the adaptive allele had reached frequency 0.99 in the population. Panel A: Mean and variance of estimates obtained from the program \texttt{sweepfinder} when applied to samples of depth 100 for a locus of size $2s/\rho$ as a function of different levels of ancestral diversity. Note that the length of the simulated locus should provide ample surrounding neutral sequence for \texttt{sweepfinder}, given that the dip in diversity is expected to be only of size $0.1s/\rho$. Panel B: Mean and variance of the estimates from our estimator (Equation 9 in the main manuscript) when applied to samples of depth 1000 for a locus of size $0.1s/(\mu+\rho)$ as a function of different mutation rates. Analogously to Figure 3 in the main manuscript we used a cutoff of $\is=5$ for the analysis. Recombination rate was always $\rho=10^{-8}$. \texttt{Sweepfinder} performs well only if the ancestral diversity is in the range $0.01$ and selection coefficients exceed $s=0.001$. Our method, in contrast, obtains reliable estimates regardless of the ancestral diversity and also for weaker selection coefficients. Note that the two methods were applied to different data sets (deep population samples of a short locus for our estimator vs.~a longer locus at only moderate coverage for \texttt{sweepfinder}). The total amount of sequence provided to either method, however, was comparable.}
 \label{fig:supp_sweepfinder}
\end{center}
\end{figure*}

\end{document}